\documentclass{emulateapj}  

\newcommand{\kms}{${\rm km~s^{\scriptscriptstyle -1}}$}
\newcommand{\kmss}{$~{\rm km~s^{\scriptscriptstyle -1}}$}
\newcommand{\kmsG}{$~{\rm km~s^{\scriptscriptstyle -1}~G^{\scriptscriptstyle -1}}$}
\newcommand{\kmsmG}{$~{\rm km~s^{\scriptscriptstyle -1}~mG^{\scriptscriptstyle -1}}$}

\newcommand{\ccm}{$~{\rm cm^{\scriptscriptstyle -3}}$}

\newcommand{\HzmG}{$~{\rm Hz~mG^{\scriptscriptstyle -1}}$}

\newcommand{\ergs}{$~{\rm erg~s^{\scriptscriptstyle -1}}$}
\slugcomment{To appear in the Astrophysical Journal}

\shorttitle{The Magnetic Field in NGC~4258}
\shortauthors{Modjaz et al.}

\defcitealias{fiebig89}{FG89}
\defcitealias{vlemmings01}{V01}
\defcitealias{vlemmings02}{V02}
\defcitealias{nedoluha92}{NW92}
\begin{document}

\title{Probing the Magnetic Field at Sub-Parsec Radii in the Accretion Disk of NGC~4258}
\author{Maryam Modjaz}
\author{James M. Moran}
\author{Paul T. Kondratko}
\author{Lincoln J. Greenhill}
\affil{Harvard-Smithsonian Center for Astrophysics, 60 Garden Street, Cambridge, MA 02138}
\email{mmodjaz@cfa.harvard.edu}

\begin{abstract}
We present an analysis of polarimetric observations at 22 GHz of the water vapor masers in NGC~4258 obtained with the VLA and the GBT. We do not detect any circular polarization in the spectrum indicative of Zeeman-induced splitting of the maser lines of water, a non-paramagnetic molecule. We have improved the 1-$\sigma$ upper limit estimate on the toroidal component of the magnetic field in the circumnuclear disk of NGC~4258 at a radius of 0.2 pc from 300 mG to 90 mG. We have developed a new method for the analysis of spectra with blended features and derive a 1-$\sigma$ upper limit of 30 mG on the radial component of the magnetic field at a radius of 0.14 pc. Assuming thermal and magnetic pressure balance, we estimate an upper limit on the mass accretion rate of $\sim10^{-3.7}~M_{\sun}$yr$^{-1}$ for a total magnetic field of less than 130 mG. We discuss the ramifications of our results on current maser models proposed to explain the observed maser emission structure and the consequences for current accretion theories. We find from our magnetic field limits that the thin-disk model and the jet-disk model are better candidates for accounting for the extremely low-luminosity nature of NGC~4258, than models that include advection-dominated accretion flows.

\end{abstract}

\section{INTRODUCTION}\label{intro_sec}

Magnetic fields appear to be important for many aspects of material-inflow and -outflow in active galactic nuclei (AGN), which are powered by accretion onto the central super-massive black holes (SMBHs). Magnetic fields are thought to provide outward angular momentum transport in the accretion process via the magneto-rotational-instability (MRI), with accompanied dissipative heating, allowing further mass-inflow \citep{balbus91, balbus03}. Strong poloidal magnetic fields may thread centrifugal winds, which could carry away angular momentum, while confining and collimating large-scale outflows and jets seen in radio-loud AGNs \citep{blandford82,krasnopolsky99}. Furthermore, broad-line region (BLR) clouds may be constrained by magnetic pressure \citep{rees87} or accelerated in magnetized winds (e.g., \citealt*{emmering92}; \citealt{bottorff97_5548}).

NGC~4258 is a very good laboratory in which to attempt to measure magnetic fields, since the magnetic field in an accretion disk could be directly measured very close to the central black hole. NGC~4258 (M~106) is a low-luminosity, weakly active Seyfert II galaxy (or of Seyfert type 1.9, as defined by \citealt{ho97_halpha}). It possesses an obscured X-ray center with an absorbing column of $N_H \sim 10^{23}$ cm$^{-2}$ and an absorption corrected $2-10$ keV luminosity of $\sim 10^{41}$\ergs~(\citealt{fruscione05}). Nuclear continuum and narrow emission lines have been detected in polarized optical light \citep{wilkes95,barth99}.

\begin{figure*}[!ht]  
\epsscale{0.35} 
\plotone{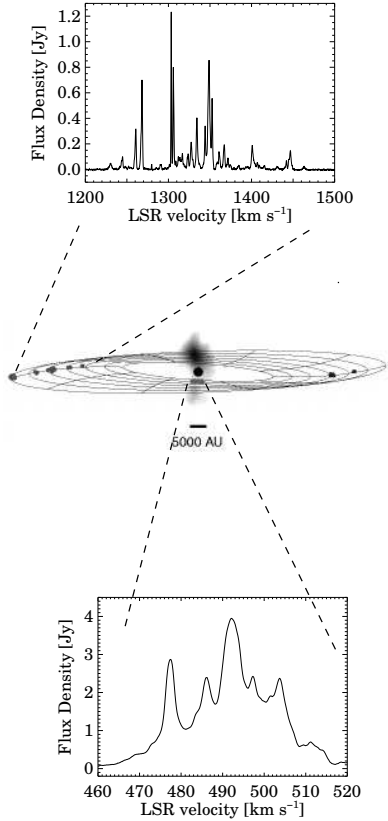}
\caption{Correspondence between maser features in velocity at 22.2 GHz and their location in the disk. \it{Middle:}\rm~The warped accretion disk, defined by the kinematics of the maser spots, is shown as a wire grid (adapted from \protect\citealt{moran99}), while the black dot marks the dynamical center of the system (the position of the SMBH). The gray-scale elongation in the center shows the 22 GHz synchrotron continuum emission \protect\citep{herrnstein98_cont}. \it{Top, bottom:}\rm~The GBT spectra were taken on 2003 October 23 (see \S~\ref{obs_sec} for details).}
\label{warpdiskfig}
\end{figure*}

The discovery of a nuclear megamaser in a survey by \citet{claussen84} in NGC~4258 and the subsequent radio single-dish and interferometry imaging studies (\citealt{greenhill95,miyoshi95}) have improved our knowledge of the nuclear region and its kinematics on sub-parsec radii. The spatial distribution and line-of-sight velocities of the masers delineate the molecular portion of a thin, Keplerian (better to 1\%) accretion disk that is nearly edge-on (tipped down by $\sim 8$\degr) and slightly warped (\citealt{moran95,herrnstein96}). The maser lines fall into three groups: a dense cluster near the systemic velocity of the galaxy, which lies close to the line-of-sight to the black hole (hereafter referred to as ``systemic maser lines''), bracketed by two groups of less densely packed lines offset by $\pm$ (800$-$1000) km~s$^{-1}$. These high-velocity masers have very small accelerations compared with the systemic masers \citep{bragg00} and are distributed over different radii close to a line perpendicular to the line of sight, the so-called ``midline'' (Fig.~\ref{warpdiskfig}); hence their line-of-sight velocities are close to the Keplerian orbital velocities.  The inner radius of the maser annulus is approximately 0.14 pc and its outer radius approximately 0.28 pc (for a distance of $D$ = 7.2 Mpc, Herrnstein et al. 1999; 1\arcsec~corresponds to 35 pc at that distance) from the central black hole with $M_{\rm{BH}} = (3.9 \pm 0.4)\times 10^7 M_{\sun}$ \citep{miyoshi95}. There is a prominent north-south twisted jet on kiloparsec scales observed in the radio, in H$\alpha$, and in X-rays (e.g., \citealt{cecil00,wilson01} and references therein). It appears to be an extension of the radio nuclear jet seen at sub-parsec distances from the dynamical center \citep{herrnstein97_jet} and to have an axis that  is well aligned with the rotation axis of the maser accretion disk. We note that the angular momentum vectors are quite misaligned at kiloparsec and sub-parsec distances from the central engine: the edge-on accretion disk, as traced by the masers, makes an angle of $\sim120\degr$ to the galactic disk of the host galaxy, a SAB(s)bc spiral \citep{tully88}, i.e., the maser accretion disk is counter-rotating with respect to the larger galactic structure.

There have been estimates made of magnetic field strengths at kiloparsec radii in NGC~4258. \citet{krause04} obtained VLA 8.44 GHz ($\lambda$3.6 cm) continuum observations of the jets of NGC~4258 and detected linear polarization at a level of $35\% - 65\%$ along the northern jets whose interpretation in the context of synchrotron emission mechanisms suggests the presence of an ordered magnetic field. Assuming energy equipartition between the magnetic field and the relativistic particles in the northern jet, and an electron-proton (electron-positron) composition, they obtained an average $B$ field of $\sim$310 $\mu$G ($\sim$90 $\mu$G). These values reflect the poloidal component of the global magnetic field at a distance of $\sim$ $1 - 4$ kpc along the jet. 

By searching for Zeeman-induced circular polarization in the maser lines, the strength and direction of the magnetic field parallel to the line of sight across the maser emission region can be directly determined. The effort to scrutinize a well-isolated, 1.2\kmss~wide maser line in the redshifted group at 1306\kmss~by \citet{herrnstein98_pol} yielded no detection of polarization in the VLA data taken in 1995. The resulting limit to the circular polarization of 0.5\% corresponds to an upper bound of approximately 300 mG for the toroidal component of the magnetic field. The Zeeman technique constitutes the most direct way of obtaining the magnitude and direction of the magnetic field. Magnetic field estimates have been derived from other techniques (e.g., \citealt{jones00}), but they are relatively indirect and model dependent, relying on assumptions of equipartition between thermal and magnetic energy, and adopted values of electron density and temperature at various radii. The initial attempt to measure polarization in NGC~4258 was made by \citet{deguchi95_4258} who reported an upper limit on linear polarization of 20\% for the high-velocity water maser components and of 15\% on the systemic components.

The format of this paper is as follows: In \S~\ref{theory_sec}, we briefly outline the theoretical background for understanding the Zeeman effect in the water molecule, and describe the technical details of how to measure the Zeeman effect. We present and discuss two data sets in \S~\ref{obs_sec} and report the measured upper limits on the magnetic field for the various maser lines in \S~\ref{results_sec}. We discuss the interpretation of the Zeeman observations in \S~\ref{discussion_sec} and give conclusions with \S~\ref{conclusions_sec}.

\section{THE ZEEMAN EFFECT IN WATER}\label{theory_sec}

	The maser emission at 22.235 GHz involves the rotational energy states $J_{\rm{K+K-}}$ = 5$_{23}$ and $J_{\rm{K+K-}}$ = 6$_{16}$ of ortho-water, which lie $\sim$600 K above the ground state. This rotational transition consists of six hyperfine components, but it is generally believed that only the strongest three hyperfine components ($F = 7-6, 6-5, 5-4$) contribute significantly to the maser emission (\citealt{deguchi86,nedoluha91,vlemmings01}, hereafter \citetalias{vlemmings01}) with velocity separations of 0.45 and 0.58\kmss~between them (those separations are marginally within a typical maser line width of $\sim0.8-$1\kmss).

\begin{figure*}[!ht]
\epsscale{1.} 
\plotone{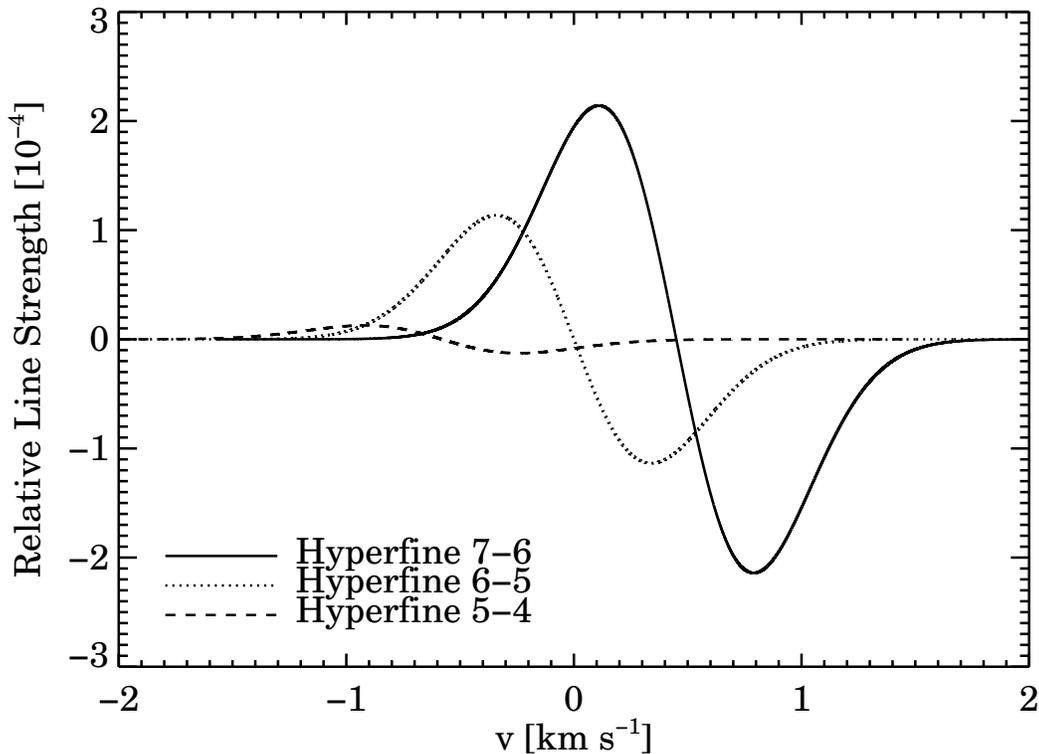}
\caption{Theoretical Stokes $V$ spectra for the three strongest hyperfine components $F = 7-6, 6-5$ and $5-4$, under the assumption of LTE at the predicted hyperfine rest velocities (centered at the hyperfine rest velocity of $F = 6-5$). The $B$ field is 50 mG and the line width 0.8\kmss~(FWHM). 
}
\label{scurvefig} 
\end{figure*}

In the presence of an external magnetic field, each hyperfine component splits into three groups of lines with characteristic polarization properties: the $\pi$ transitions ($\Delta M_F = 0$) produce radiation linearly polarized along the external $B$ field, with negligible frequency displacements from the parent-hyperfine transition. The $\sigma^{\pm}$ transitions ($\Delta M_F = \pm 1$) generate emission whose $E$ field vector rotates in a circle in the plane perpendicular to the external $B$ field direction and are symmetrically offset from the parent-hyperfine in frequency by $\pm\Delta \nu_Z$ (we follow the notation used in \citetalias{vlemmings01}; see their Fig. 1 showing the theoretical Zeeman pattern). Since the water molecule is non-paramagnetic, the splitting is very small ($\Delta \nu_Z \sim 1$ Hz mG$^{-1}$ or $10^{-5}$\kmsmG~compared to, e.g. up to 3 kHz mG$^{-1}$ or 0.6\kmsmG~for the OH molecule at 18 cm) and much less than the thermal line width of the maser features for expected sub-Gauss $B$ fields ($\Delta \nu_Z \sim 10^{-4}-10^{-3} \Delta \nu_L$). For a $B$ field along the line of sight, the $\sigma^{\pm}$ components appear in left-circular polarization (LCP) and right-circular polarization (RCP), respectively, and their slight frequency separation can be detected as an {\sl ``S''} shaped profile in the Stokes $V$ [= (LCP $-$ RCP)/2~] spectrum for a simple Gaussian line profile (Fig.~\ref{scurvefig}). This {\sl S} profile is proportional to the derivative of the Stokes $I$ [=(LCP + RCP)/2~] spectrum in the small splitting case (\citealt{fiebig89}, hereafter \citetalias{fiebig89}; \citealt{herrnstein98_pol}; \citetalias{vlemmings01}; see also Appendix~\ref{thermalz_sec}). 

By modeling the theoretical H$_2$O Zeeman pattern and convolving it with Gaussian profiles of differing line widths, \citetalias{fiebig89} constructed a set of synthetic Stokes $V$ spectra and found the following relationship between the applied $B$ field and the resulting circular polarization:

\begin{equation}
                \frac{V_{\rm{max}}}{  I_{\rm{max}}} = \frac{A_{\rm{F-F'}} B_{||}} {\Delta v_{L}} , 
\label{zeeman_eq}
\end{equation}
where $V_{\rm{max}}$ is the maximum of the $S$ curve, $I_{\rm{max}}$ is the maximum in Stokes $I$, $B_{||}$ is the magnetic field parallel to the line-of-sight, $B_{||} = B\rm{cos}\theta$ (where $\theta$ is the angle between the $B$ direction and the line of sight), and $A_{\rm{F-F'}}$ is a coefficient that depends on which hyperfine component is assumed. In the case where the hyperfine lines appear individually, $A_{7-6}$ = 0.0134\kmsG~for the $F=7-6$ hyperfine that possesses the largest transition probability; $A_{6-5}$ = 0.008\kmsG~for $F=6-5$ and $A_{5-4}$ = 0.001\kmsG~for $F = 5-4$.

Equation~(\ref{zeeman_eq}) does not account for the non-thermal nature of maser emission, which includes inverted, non-thermal, population distributions and amplification of emission. \citealt{nedoluha92} (hereafter \citetalias{nedoluha92}), \citealt{watson01} and, more recently, \citealt{vlemmings02} (hereafter \citetalias{vlemmings02}), conducted a series of studies in which they solved the radiative transfer and rate equations numerically for water masers in a magnetic field, extending the pioneering analytic work of \citet*{goldreich73}. These numerical simulations account for saturation effects and the presence of the three strongest hyperfine components and their magnetic substates. Their major findings were that the ``thermal'' equation~(\ref{zeeman_eq}) is an excellent approximation for the case of maser polarization induced by a magnetic field with an effective coefficient $A_{\rm{F-F'}}$ of 0.020\kmsG~(appropriate for the presence of the three strongest hyperfines). The estimate of the $B$ field via equation~(\ref{zeeman_eq}) is accurate to within a factor of 2, as long as the water maser is not highly saturated, i.e., has a linewidth $\la$ 0.8 \kms \citepalias{nedoluha92}. In those numerical simulations, maser propagation was modeled in one dimension. However, initial calculations in 2D by \citetalias{vlemmings02} suggest that the resulting Zeeman $S$ curves might be slightly narrower and have a distorted profile, than dedyced from 1D calculations. \citet{gray03} did a study of maser polarization models in which he compared his general models to the ones put forth by Watson and collaborators \citep{watson94,nedoluha94} and found that his equations can be reduced to the ones by Watson, for the limiting case of small Zeeman splitting.

In summary, we adopt $A_{\rm{F-F'}}$ = 0.020 km~s$^{-1}$~G$^{-1}$ for the $S$-curve model, where the Stokes $V$ profile is proportional to the derivative of Stokes $I$, appropriate for the presence of the three strongest water hyperfines in a masing environment. This Zeeman splitting coefficient is larger by a factor of 1.5 than for the LTE model where solely the hyperfine component with the largest transition probability under LTE is present. Following other Zeeman experiments in the radio regime (e.g., \citealt{goodmanphd}, \citealt{heiles01}), we adopt the IEEE convention for circular polarization, where a magnetic field pointing away from the observer is $positive$ and produces a LCP component at a $lower$ velocity than the RCP component.

Several water maser polarization experiments have been conducted and interpreted in the framework of equation~(\ref{zeeman_eq}), with line-of-sight magnetic fields inferred on the order of $10-50$ mG in Galactic star-forming regions (\citetalias{fiebig89}; \citealt{sarma01,sarma02}). These results agree with magnetic field strengths expected for a wide range of probed densities in molecular clouds, which are subject to flux freezing and ambipolar diffusion, thus providing corroborating evidence for the cogency of the Zeeman interpretation. Furthermore, line-of-sight magnetic fields have been inferred on the order of $50-3000$ mG in circumstellar envelopes of late-type stars \citepalias{vlemmings02}. These values are consistent with $B$ field estimates based on OH maser observations for a reasonable radial scaling of the $B$ field.

There are two types of maser lines in the spectrum of NGC~4258, which require slightly different treatments to measure the Zeeman effect. For the well-isolated redshifted and blueshifted high-velocity maser lines, the standard $S$-curve analysis method as outlined above works well. For the systemic features, which are heavily blended in velocity space, we developed a variant of the $S$-curve method. Our new technique is more appropriate since it does not require the a priori knowledge of the velocity components. In this procedure, we cross correlate the LCP and RCP spectra to determine the net velocity offset (see Appendix~\ref{crosscor_subsec} for details). This cross correlation method (see also \citealt{heiles88} for a slightly different treatment in the case of observations of blended thermal \ion{H}{1} features) is based on the assumption that the $B$ field strength and direction remain constant over the emission region, i.e., that the Zeeman-induced velocity shift is constant across the spectrum. 

From simple statistical arguments, supported by Monte Carlo simulations (Appendix~\ref{montecarlo_subsec}), the rms sensitivity to the LOS $B$ field ($\sigma_{B_{||}}$) is given by the following relationship (which applies for both the $S$-curve and the cross correlation method), \\
\begin{equation}
\sigma_{B_{||}} = \frac{1}{2}\frac{\Delta v_L}{A_{\rm{F-F'}}~SNR}\frac{1}{\sqrt{N}}~~,
\label{sigmab_eq}
\end{equation}
where $\Delta v_L$ is the FWHM line width, $A_{\rm{F-F'}}$ the Zeeman splitting coefficient, $SNR$ the signal-to-noise ratio for the specific line (in Stokes $I$) and $N$ the number of lines over which the fit is performed. This result is appropriate if the $N$ lines are not heavily blended. For differing line strengths and line widths of the spectrally blended lines, as found in the cross correlation technique, the $\Delta v_L /(SNR~\sqrt{N})$ factor must be replaced by an appropriate average.

\section{OBSERVATIONS AND DATA REDUCTION}\label{obs_sec}

\subsection{VLA Data}\label{vlaobs_sec}
  
Data were taken on 2001, December 16 in spectral line and full polarization mode with the Very Large Array (VLA) of the National Radio Astronomy Observatory (NRAO)\footnote{The National Radio Astronomy Observatory is operated by Associated Universities, Inc., under the cooperative agreement with the National Science Foundation.} in its D configuration for approximately 12 hours at K Band (22.2 GHz). The resulting synthesized half-power beamwidth is 2.8 arcsec. In this beam, the maser spots (with intrinsic sizes of less than 100 $\mu$arcsec and inter-maser-separations of several milliarcsec as shown by VLBI observations; \citealt{miyoshi95,moran95}) appear as one point source and are not resolved; therefore, beam effects such as beam squint and beam squash do not affect our data. The segregation among the various maser spots occurs in velocity space, since the maser spots possess different orbital velocities. The observations of an isolated high-velocity feature at 1269~km~s$^{-1}$ had to be interlaced with those of the systemic features due to the limited bandwidth of the VLA. The calibrators were observed separately for the two different frequency bands. The observations were performed with a bandwidth of 1.56 MHz with 128 channels, yielding a channel spacing of 12.2 kHz (or 0.16 \kmss). With uniform weighting the spectral resolution is 0.19\kmss, which corresponds to $\sim1/5$ of the high-velocity maser line width. The first three hours of the observations experienced snow fall, with skies gradually clearing thereafter.

\begin{figure*}[!!ht]
\plotone{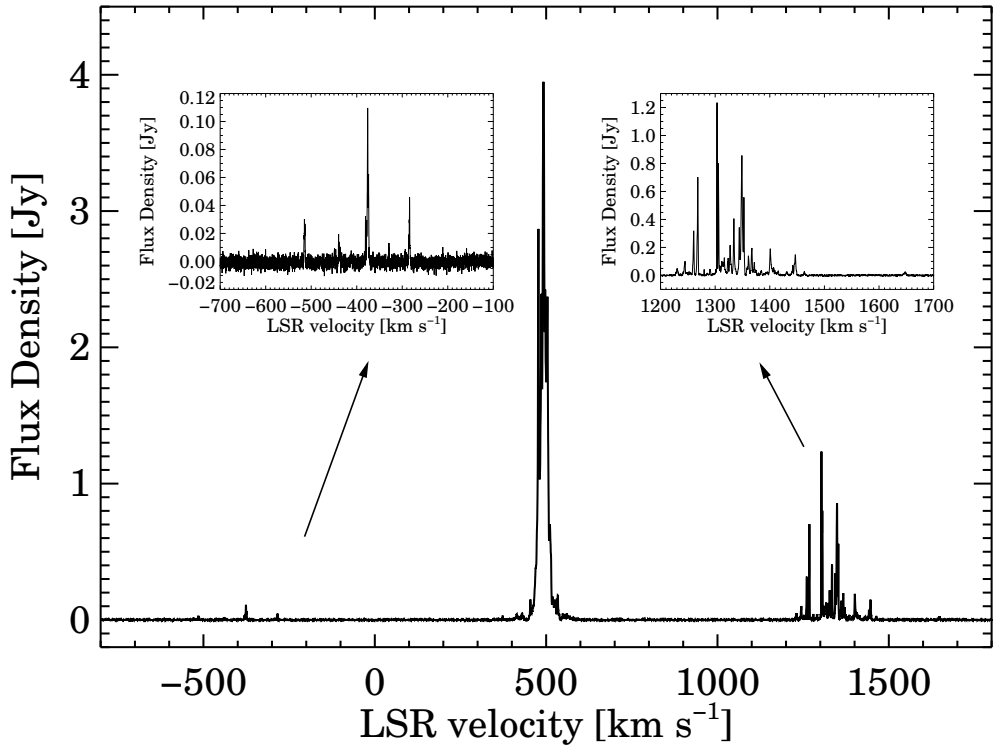}
\caption{GBT total power data with a bandwidth of 200 MHz. Note the flat baseline after removing a running mean that excluded the line-displaying regions. Inserts show zooms on the red- and blue-shifted high-velocity lines, respectively. The rms noise in each channel is 2.9~mJy. The feature at 1647\kmss~is a highly significant new discovery. It is the most redshifted high-velocity feature discovered in NGC~4258, and it extends the inner boundary of the maser disk to 0.12 pc, based on the Keplerian rotation curve.}
\label{allgbtdatafig}
\end{figure*}
 
The data were edited, reduced, and calibrated with standard spectral line techniques in AIPS\footnote{Astronomical Image Processing System of NRAO.}, including bandpass, amplitude and phase calibration. Amplitude calibration was performed with respect to 3C~147, a VLA calibrator with known flux density at 22 GHz, phase calibrations relative to 1146+399, and bandpass calibrations using observations of 0854-201. The system temperatures of the K-band receivers in nine of the antennas were in the range of $100-130$ K, whereas the new K-band receivers, residing in 17 out of the 26 operational antennas at the date of observations, had system temperatures of $60-80$ K. The amplitude calibration is estimated to be accurate to within 15\%. Strong internal radio-frequency-interference was present in the calibrator data taken with the old receivers at the frequency corresponding to that of the systemic maser emission and had to be removed from the data set. 

We performed linear polarization calibration in two steps: first we used the observations of the strong source 0854+201 over a range of parallactic angles to correct for the imperfect feed responses (as expressed by the so-called ``leakage terms'', which are on the order of a few \% for the VLA K band feeds). Secondly, we determined the phase offset between LCP and RCP by observing 3C286, a polarized source with a known and invariant polarization angle. For circular polarization calibration, an additional step involved correcting for the differences in gain for LCP and RCP and was performed during the analysis of the polarization data (see \S~\ref{results_sec}). 
\begin{deluxetable*}{lccc}
\tablecaption{Journal of H$_2$O Maser Observations}
\tablehead{
\colhead{Parameter} & \multicolumn{2}{c}{VLA Data\tablenotemark{\dag}}& \colhead{GBT Data\tablenotemark{\dag\dag}} \\
\cline{2-3} \\
 & Set I\tablenotemark  & Set II\tablenotemark & \\
}
\startdata
Date 	                & 2001 Dec 16 	& 2001 Dec 16 	& 2003 Oct 23 	\\
Radio LSR center velocity\tablenotemark{a} [\kms]	& 495		& 1268 & 	550	\\
Total Bandwidth [\kms] & 21	& 21		& 2700  	\\
Channel spacing	[\kms]	& 0.16		& 0.16		& 0.16		\\
Integration time\tablenotemark{b} [min]& 241 	& 231		& 348		\\
1 $\sigma$ noise in Stokes $I$\tablenotemark{c} [mJy] & 2.4\tablenotemark{d}	& 2.4	    &      2.9 \\
\enddata
\tablenotetext{\dag}{RR, LL, RL, LR observed.}
\tablenotetext{\dag\dag}{RR and LL observed.}
\tablenotetext{a}{V$_{\rm{hel}} = V_{\rm{LSR}} -29.169$ \kmss.}
\tablenotetext{b}{Effective time spent on source, not including time spent on calibrators.}
\tablenotetext{c}{Noise per channel, corrected for variable antenna gain and referenced to outside the atmosphere.}
\tablenotetext{d}{Derived from line-free intervals at high velocities.}
\label{datatab}
\end{deluxetable*}

\subsection{GBT Data}\label{gbtobs_sec}

We took spectral line data in RCP and LCP on 2003 October 23 with the Robert C. Byrd Green Bank Telescope (GBT) of the NRAO. With an off-axis feed arm, the GBT delivered an unblocked aperture and low sidelobes with a FWHM beamwidth of 33\arcsec~at 22.2 GHz. The spectrometer had a bandwidth of 200 MHz configured to provide a channel spacing of 0.164\kmss~with its 16384 channels over 200 MHz (2700 \kms). The K-band receiver for the upper frequency range (22$-$26.5 GHz) had two beams, each beam producing two polarization channels. The data were taken in beam-switched mode with beam throws of the order of 4\arcmin, the angular distance between the pair of beams. The cycle time of two minutes was sufficient to compensate for atmospheric fluctuations. In this fashion, one beam of the telescope was always pointing at the source. A total of eight hours of data were taken, including pointing and focusing observations after every hour of source observation. Applying the local pointing corrections improved the pointing uncertainty to $\la$ 2$\arcsec$. The effective on-source time was 348 minutes and the zenith opacity $\sim$0.07.

The gain curve, which describes the efficiency of the antenna as a function of elevation, had been determined from observations of the radio source NGC~7027 over a range of elevations. The gain curve with the active surface-control enabled indicates a peak antenna efficiency of $\eta$ = 0.6 at an elevation of $\sim 30\arcdeg$ and stays comparatively flat ($\eta\approx 0.5-0.6$) over a wide range of elevations ($20\arcdeg-80 \arcdeg$). The rms deviation of the gain measurements around the best-fit polynomial model, which is applied to the data for gain correction, was about 10\%. Amplitude calibration was performed via injection of a signal from a noise diode every 0.25 sec. Data reduction was performed with a customized IDL procedure.

The residual baseline variations in the spectra were eliminated by subtracting a running mean computed over intervals of 256 channels (41\kmss) that excluded the line-bearing regions; this procedure yielded flat baselines in the final spectrum (Fig.~\ref{allgbtdatafig}). The system temperatures for the GBT data ranged from $35-50$ K and quoted flux densities are accurate to 10\%. The largest source of uncertainty is the inaccurate knowledge of the strength of the noise diodes over the large bandwidth of the GBT data. We note that the uncertainty in the absolute intensity scale has little impact onto the reported results of the magnetic field upper limits: only a differential measurement, the SNR, determines the upper limit (eq.~[\ref{sigmab_eq}]).

Table~\ref{datatab} lists a summary of the specifications of both data sets. The velocities in this paper follow the radio definition of Doppler shift ($\Delta\nu / \nu = - v / c$) and are computed with respect to the local standard of rest (LSR).

\section{RESULTS AND INTERPRETATION}\label{results_sec}
\begin{figure*}[!ht]
\epsscale{0.5}
\plotone{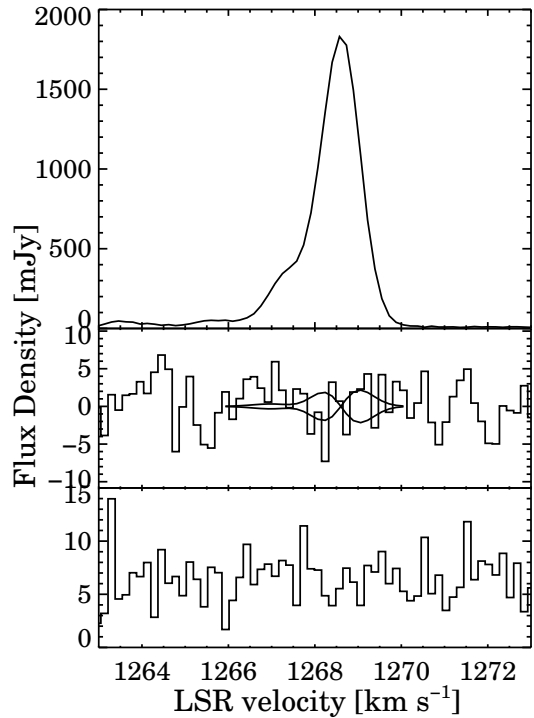}
\caption{VLA data of the high-velocity feature at 1269~\kms~with Zeeman analysis. \it{Top:}\rm~Stokes $I [=(\rm{LCP} + \rm{RCP})/2 ]$ spectrum. \it{Middle:}\rm~The Stokes $V [=(\rm{LCP} - \rm{RCP})/2 ]$ spectrum, which is adjusted by having removed a scaled replica of the total power profile, determined by the $c_2$ parameter in equation~(\ref{stokesv_eq}) for this fit. Superimposed are the Zeeman $S$ curves for the upper limit of $\pm 90$ mG on the LOS magnetic field as reported in Table~\ref{zeemanfittab}. \it{Bottom:}\rm~Total linear polarization spectrum [Stokes $(Q^2 + U^2)^{1/2}$].}
\label{vlahighvfig}
\end{figure*}

\begin{figure*}[!ht]
\epsscale{0.5}
\plotone{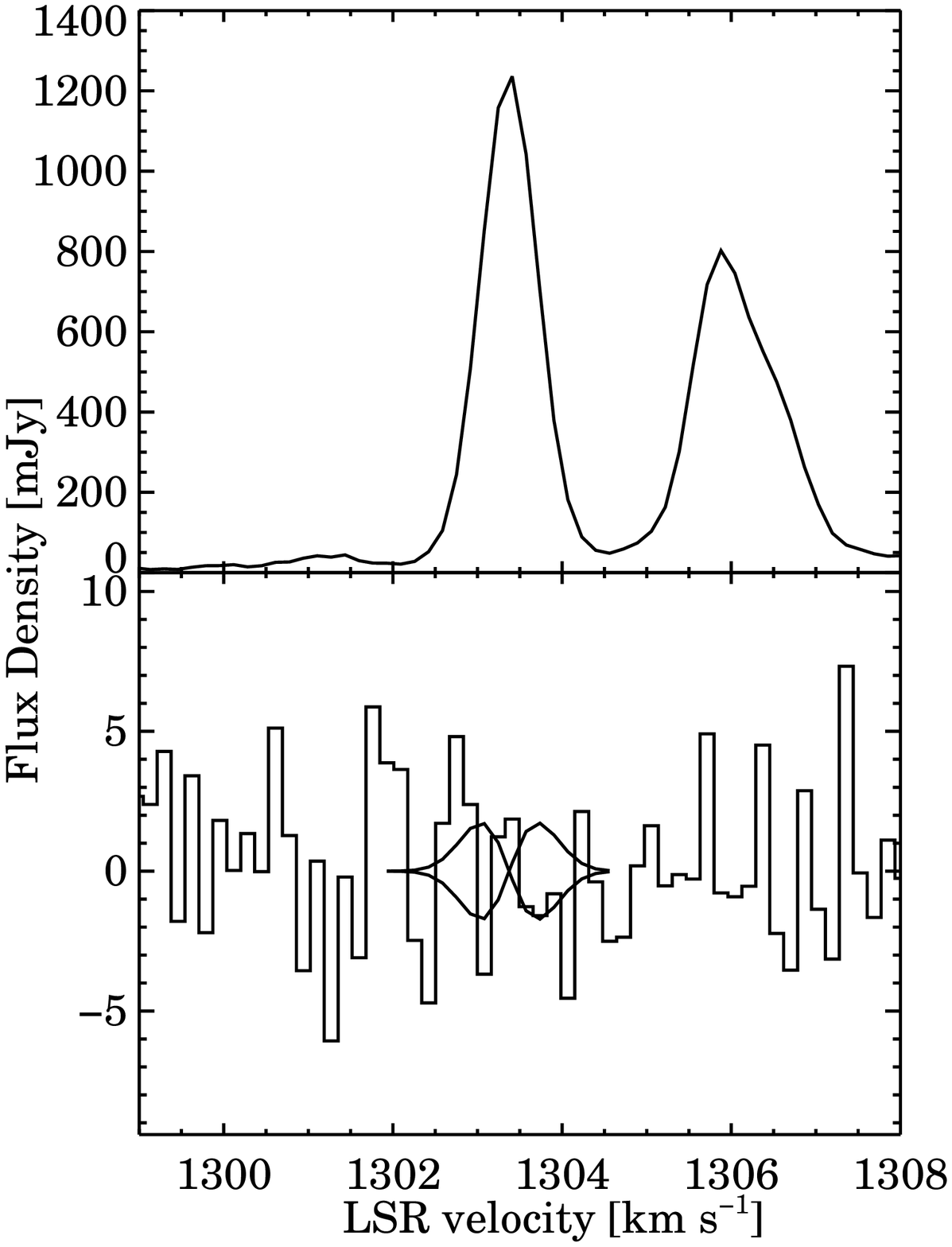}
\caption{Similar to Figure~\ref{vlahighvfig}, but for the strongest redshifted high-velocity feature in the GBT spectrum, except no linear polarization data could be obtained with the GBT spectrometer. The 1-$\sigma$ $S$ curves for $B^{\rm{lim}}_{||} = \pm$90 mG is overplotted.}
\label{gbthighvfig}
\end{figure*}
For circular polarization data, instrumental gain differences between LCP and RCP cause a portion of the Stokes $I$ profile to appear in the Stokes $V$ spectrum. This effect is mitigated by simultaneously fitting a replica of Stokes $I$ to the Stokes $V$ spectrum in addition to a scaled derivative of Stokes $I$ in the least-squares sense (e.g., \citealt{troland82,sault90}) : 
\begin{equation} 
V(v) = c_1~\frac{dI(v)}{dv} + c_2~I(v) .
\label{stokesv_eq}
\end{equation}
\noindent Here, $c_2$ denotes the instrumental polarization term. The factor $c_1$ is a measure of the Zeeman splitting and corresponds to $ A_{\rm{F-F'}} B \rm{cos}\theta / \sqrt{2}$ (Appendix~\ref{thermalz_sec}). This calibration procedure is equivalent to the assumption that equal flux should be present in both the LCP and RCP spectra, i.e., that the circular polarization spectra are intrinsically anti-symmetric. However, theoretical maser polarization simulations (see \S~\ref{theory_sec}) produce $V$ spectra that are asymmetric due to the presence of hyperfine structure. Efforts to verify the intrinsic shape of the $V$ spectrum observationally are hampered by the the fact that it has been impossible to determine the instrumental circular polarization independently well enough to the requisite degree of $P_V = V / I \sim0.1\%$ \citep*{zhao93}. In order for the derived magnetic fields strength $B_{||}$ to be counted as a detection, we require the field strength to be greater than 3 times the associated uncertainty, $\sigma_{B_{||}}$. With this criterion, we did not make any Zeeman-induced circular polarization detection. 



\subsection{High-velocity Features}\label{polresults_highv_sec}

Figure~\ref{vlahighvfig} shows the Stokes $I [= (\rm{LCP} + \rm{RCP})/2]$, i.e., total power, VLA spectrum of the redshifted high-velocity feature at 1269~km~s$^{-1}$ with a thermal rms noise of about 2.4 mJy (in a 0.16\kmss~resolution interval; see Table~\ref{datatab}). The high-velocity line was best fit by a two-component Gaussian, such that the main line is centered at 1268.6\kmss and has a FWHM line width of $\Delta v_L$ = 1.01\kmss. 
A formal minimum $\chi^2$ fit of the Stokes $V [=(\rm{LCP} - \rm{RCP})/2]$ spectrum to equation~(\ref{stokesv_eq}) indicated a 1-$\sigma$ upper limit to any intrinsic circular polarization of $P_V \la 0.08$\% in the 1269~km~s$^{-1}$ feature. For a maser Zeeman splitting coefficient of $A_{\rm{F-F'}} = 0.020$\HzmG, $\sigma_{B_{||}}\approx$ 40 mG (1 $\sigma$). This value is the same, as expected, as the value derived from equation~(\ref{sigmab_eq}) and Monte Carlo Simulations (Appendix \ref{montecarlo_subsec}). From a conservative estimate for the upper limit on the magnetic field along the line-of-sight, $\mid B^{\rm{lim}}_{||}\mid = \mid B^{\rm{output}}_{||}\mid + \sigma_{B_{||}} + \sigma_{\sigma_{B_{||}}}$, we obtained $B_{||}\la$~90~mG (Table~\ref{zeemanfittab}), where $\sigma_{\sigma_{B_{||}}}$ stands for the uncertainty in $\sigma_{B_{||}}$ and is estimated from Monte Carlo simulations (\S~\ref{montecarlo_subsec}). This feature is located at a radial distance of 0.27 pc from the SMBH (70,000 in units of Schwarzschild radii, $R_S = 2 GM_{\rm{BH}}/c^2$, for $M_{\rm{BH}} = 3.9 \times 10^7 M_{\sun}$). Following convention, the middle panel of Figure~\ref{vlahighvfig} shows the Stokes $V$ spectrum intrinsic to the source, i.e., after removing a residual replica of Stokes $I$. Superimposed are the 1-$\sigma$ Zeeman $S$ curves for B$_{||}^{\rm{lim}} = \pm$90 mG. The bottom panel of Figure ~\ref{vlahighvfig} shows the total linear polarization [Stokes $(Q^2 + U^2)^{1/2}$] spectrum: no fractional linear polarization [Stokes $(Q^2 + U^2)^{1/2} / I$] at a level of 0.1\% (1 $\sigma$) is detected. 

	For the GBT dataset, the combination of low system temperatures and a new spectrometer with a large bandwidth promised higher sensitivity for Zeeman detection. However, the variable nature of masers caused a decrease in the SNR during our observations and no Zeeman-induced circular polarization could be detected in any of the numerous features. The $S$-curve analysis yielded an upper limit for the strongest high-velocity feature at 1303.3\kmss~of B$_{||}\la$90 mG, similar to the VLA experiment (Table~\ref{zeemanfittab}). Figure~\ref{gbthighvfig} shows the Stokes $I$ spectrum (top) with a thermal rms noise of 2.9 mJy (in a 0.16\kmss~resolution interval). This feature is best-fit by a one-component Gaussian model of width 0.79\kmss~(FWHM), centered at 1303.3\kmss. The instrumentation-corrected Stokes $V$ spectrum is displayed in the bottom panel of Figure~\ref{gbthighvfig} with the overlayed 1-$\sigma$ $S$ curves for B$_{||}^{\rm{lim}} = \pm 90~\mathrm{mG}$. This feature is located at a distance of 0.25 pc from the central black hole, which corresponds to 62,000 $R_S$. Linear polarization data could not be obtained with the GBT spectrometer at the time of observations.

In this case, B$_{||}$ corresponds to the toroidal magnetic field and our limits are significant on the assumption that B$_{||}$ is not tangled over the maser emission region of $\la$~0.003 pc (or $\la 10^{16}$ cm) \citep{moran95}. If the 22.2 GHz water vapor line consists solely of the $F = 7-6$ hyperfine component, the $B$ field estimates have to be multiplied by a factor of 1.5. This result is an improvement of the upper limit of B$_{||}\la$~300 mG derived by \citet{herrnstein98_pol} on a feature at 1306~km~s$^{-1}$ using a similar method.

Hydro dynamic and MHD turbulence are expected in accretion disks. Assuming a Kolmogorov power spectrum for turbulent magnetic fields in a circumnuclear disk, \citet{watson01} produced realistic simulations of total-power water maser spectra and found that the consequences for circular polarization were not serious. In the case of turbulent fields, they suggested that equation~(\ref{zeeman_eq}) underestimates the magnitude of the magnetic field generally by a factor of about 3. In their calculations, they assumed that the turbulence length for the radial scale is the same as the scale height of the disk, $H$, as they assumed isotropic turbulence where the longest wavelength is set by the height. However, C. Gammie (2004, private communication) estimated that most of the power may be contained on scales of the order 40$H$. Since the length of the masers and $H/R$ are estimated to be $\la$0.003 pc and $\la$0.0025, respectively \citep{moran95}, the turbulence length according to Gammie is larger than the maser length, by a factor of 4$-$10. For turbulence lengths that are considerably larger than the maser amplification length, few magnetic field reversals are expected across the maser length. Thus in summary, we conclude that it is unlikely that the non-detection of Zeeman-induced circular polarization is due to a scenario where turbulent  magnetic fields quench the maser polarization signal due to magnetic field reversals.

\subsection{Systemic Features}\label{polresults_sys_sec}

\begin{figure*}[!ht]
\epsscale{0.5}
\plotone{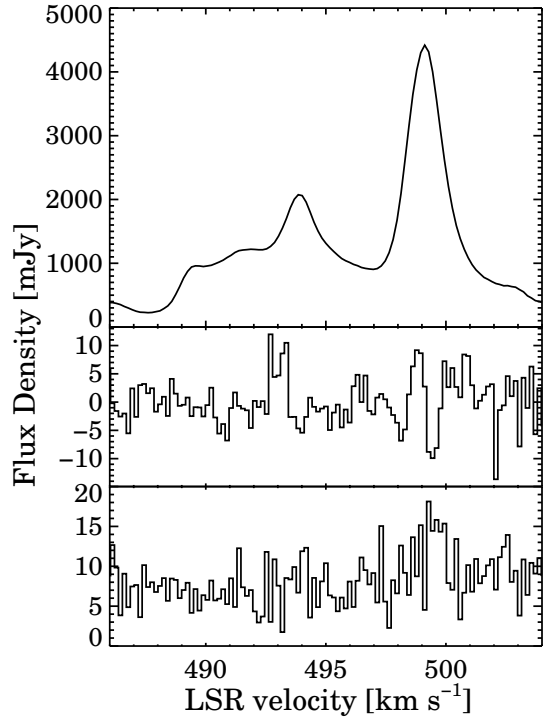}
\caption{\it{Top:}\rm~VLA total power [Stokes $I$ = (RCP+LCP)/2] data of the systemic features.
\it{Middle:}\rm~Renormalized circular polarization spectrum [Stokes $V$ = (RCP$-$LCP)/2]. \emph{Bottom:} Linear polarization data [Stokes $(Q^2 + U^2)^{1/2}$].}
\label{vlasysfig}
\end{figure*}
\begin{deluxetable*}{cccccccccc}
\tablecaption{H$_2$O Maser Parameters and Zeeman Interpretation}
\tablehead{
\colhead{Data Set} & \colhead{Line center\tablenotemark{a}} & \colhead{FWHM} & \colhead{I$_{\rm{o}}$} & \colhead{Fitted Range}&
\colhead{B$_{||}$\tablenotemark{b}} &\colhead{$\sigma_{B_{||}}$\tablenotemark{b}}&\colhead{$\sigma_{\sigma_{B_{||}}}$\tablenotemark{c}} & \colhead{B$^{\rm{lim}}_{||}$ (1$\sigma$)\tablenotemark{d}}& \colhead{R}\\
 & [\kms] & [\kms]  & [mJy] &  [\kms]& [mG] & [mG] & [mG]& [mG]  & [pc] \\
}
\startdata \\
VLA\tablenotemark{e}	& 1268.6& 1.01 & 1795&	1266.5$-$1270.0&  -47 & 38 & 5 & 90 & 0.27\tablenotemark{g}\\
GBT\tablenotemark{e} 	& 1303.3& 0.79 & 1212&  1302.0$-$1304.5&  40 & 44 & 6 & 90 &0.25\tablenotemark{g}  \\
VLA\tablenotemark{f} 	& 499.1	& ...	&  4553  & 486.0$-$504.0 & 29	& 16 & 4& 49 & 0.14\tablenotemark{h}\\
GBT\tablenotemark{f} 	& 492.3	& ...   &  4102  & 472.0$-$507.0 & 7	& 19 & 4& 30 &0.14\tablenotemark{h}\\
\enddata
\tablenotetext{a}{Radio LSR velocity.}
\tablenotetext{b}{Values for $B_{||}$ and $\sigma_{B_{||}}$ have been obtained from the Zeeman equation~(\ref{zeeman_eq}) with $A_{\rm{F-F'}}$ = 0.020 \kmsG~\citep{nedoluha92}, which presumes that the 22.2 GHz water maser line is a merger of the three strongest hyperfine lines. The values for $\sigma_{B_{||}}$ agree with values derived from Monte Carlo simulations with characteristic noise and data parameters (see \S~\ref{montecarlo_subsec}).}
\tablenotetext{c}{Values for $\sigma_{\sigma_{B_{||}}}$ were obtained from Monte Carlo simulations with synthetic data that display the same data and noise characteristics as the corresponding datasets (see \S~\ref{montecarlo_subsec}).}
\tablenotetext{d}{Obtained from a conservative estimate for B$^{\rm{lim}}_{||}$ = $\mid B^{\rm{output}}_{||}\mid$ + $\sigma_{B_{||}}$ +$\sigma_{\sigma_{B_{||}}}$. If the 22.2 GHz water vapor line consists solely of the $F = 7-6$ hyperfine component, the $B$ field estimates would be multiplied by a factor of 1.5.}
\tablenotetext{e}{$S$ curve fits performed.}
\tablenotetext{f}{Cross correlation method performed. Also, the line parameters (line center velocity, flux density) listed are for the strongest systemic feature in the spectrum.}
\tablenotetext{g}{Radial distance of maser feature from the central black hole. Positions derived from the line-of-sight velocities of the high-velocity features, assuming a Keplerian rotation curve and the standard values for $M_{\rm{BH}}$, distance, systemic velocity of the galaxy and that the features lie on the midline. }
\tablenotetext{h}{Radial distance of maser feature from the central black hole. Position taken from \citet{bragg00}, which is based on the observed acceleration.}
\label{zeemanfittab}
\end{deluxetable*}

For the cross correlation method, we performed an equivalent gain calibration procedure by requiring the integrated line fluxes of LCP and RCP to be equal. Simulations and application of both calibration methods to the well-isolated features yield the same $B$ field estimates and uncertainties. The systemic maser lines are stronger and more numerous, thereby improving the sensitivity to the $B$ field, but have broader line widths than the high-velocity features and are heavily overlapped, decreasing the sensitivity.
\begin{figure*}[!ht]
\epsscale{0.5}
\plotone{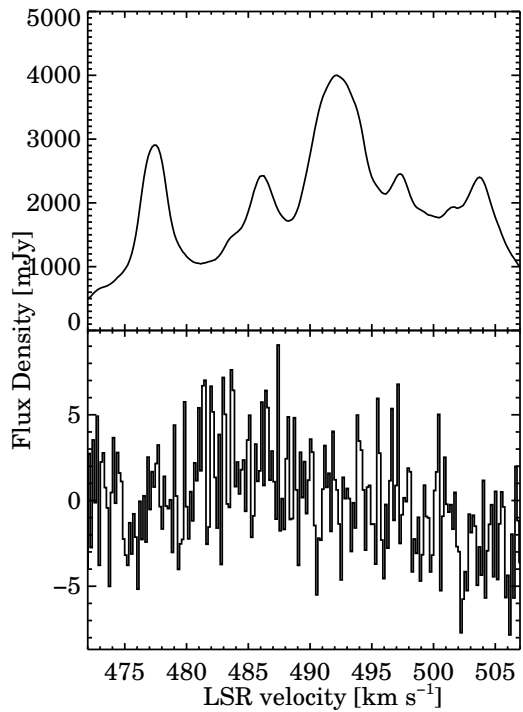}
\caption{\it{Top:}\rm~GBT total power [Stokes $I$ = (RCP+LCP)/2] data of the systemic features.
\it{Bottom:}\rm~Renormalized circular polarization spectrum [Stokes $V$ = (RCP$-$LCP)/2].}
\label{gbtsysfig}
\end{figure*}
For the VLA data, we performed the cross correlation technique on those systemic maser lines that occupy the inner 90\% of the bandwidth, in order to avoid passband problems at the edges of the spectrum. We derived a value of $B_{||} = 29 \pm 16$ mG (Fig.~\ref{vlasysfig}). Though this result is formally a 2 $\sigma$ detection, we do not count it as a significant detection (defined as a detection with a significance level of 3 $\sigma$), but rather report an upper limit of B$_{||}\la$~50 mG (see Table~\ref{zeemanfittab} for computation). There are deviations near 493\kmss~and at 499\kmss~in the Stokes $V$ spectrum, which are responsible for the 2$\sigma$ result. There is no clear detection of fractional linear polarization at a level of 0.3\% (1 $\sigma$) (Fig.~\ref{vlasysfig}, bottom panel). However, the conditions in the masing medium could give rise to internal Faraday rotation along the maser amplification path, leading to depolarization of a strong linearly polarized signal (see, e.g., \S 13.6 of \citealt{tms}). The requisite electron density for complete depolarization by this mechanism (i.e. differential Faraday rotation of $\pi$) is $n_e \approx \pi \left(8.1 \times 10^5 \lambda^2 B_{||} L \right)^{-1}$, where $\lambda$ is the observing frequency in meters, $B_{||}$ the $B$ field along the line-of-sight in Gauss, $L$ the pathlength over which depolarization is taking place in pc, and $n_e$ is the electron density in cm$^{-3}$. For $\lambda$ = 1.35 cm, $L$ = 0.003 pc \citep{moran95}, $B_{||} \la 0.03$ G, one obtains $n_e \sim $200 cm$^{-3}$. For an assumed average molecular density of $10^{9}$\ccm, this corresponds to an ionization fraction of $x_e \sim$ 10$^{-7}$. These low values are well within the range of the expected ionization fraction. Models by \citet*{neufeld94} and \citet*{neufeld95}, in which oblique X-ray illumination is heating the gas in the molecular accretion disk of NGC~4258 and other megamaser galaxies, give a theoretical upper limit of $x_e \sim 10^{-5}$. Note that there are observations of strong linear polarization (up to 65\%) in galactic water masers \citep*{garay89}, which arise in star formation regions.

Cross correlating the GBT spectra of the systemic features in LCP and RCP leads to a 1-$\sigma$ upper limit of B$_{||}\la$30 mG (Fig.~\ref{gbtsysfig} and Table~\ref{zeemanfittab}). We chose the velocity range of the systemic maser lines such as to maximize the number of included features, while minimizing the extent of the baseline modulation seen in the renormalized Stokes $V$ spectrum (bottom panel of Fig.~\ref{gbtsysfig}). Note that the different appearance of the systemic features in the two datasets are due to both time variability and centripetal acceleration of $\sim$10 \kms yr$^{-1}$ \citep{bragg00}. 

In the case of the systemic features, $B_{||}$ refers to the ordered radial component of the $B$ field, assuming the features are threaded by a $B$ field constant in magnitude and direction. They are located at a distance of $R \approx 0.14$ pc ($37,000~R_S$) from the central black hole \citep[as derived from the observed acceleration of the systemic features]{bragg00}. The features are distributed over an azimuthal angle in the disk of about 8$\arcdeg$, or about 0.02 pc. Recently, \citet{herrnstein05} estimate that the systemic features are located in a bowl-shaped depression of the warped disk and occupy a region of length 0.05 pc along the LOS.  
 
\noindent Table~\ref{zeemanfittab} lists the details and summary of the individual Zeeman fits and their interpretation.


\subsection{Saturation Effects}\label{saturation_sec}

Some theoretical investigations have shown that the degree of observed circular water maser polarization depends also on the saturation state of the masers. Masers begin to saturate when $R / \Gamma \ga 1$, and reach full saturation when $R / \Gamma \ga 100$, where $R$ is the rate of stimulated emission and $\Gamma$ the decay rate, with $\Gamma$ = 1 s$^{-1}$ from the lifetime of involved IR transitions.

The stimulated emission rate of a maser is given by $R = A~S_{\nu}~c^2/(8~\pi~h \nu^3) \times (\Omega_b /\Omega_s)$ \citep*{reid88}, where $A$ is the Einstein coefficient for spontaneous emission ($2 \times 10^{-9}$ s$^{-1}$ for the water transition), $S_{\nu}$ the specific flux density of the maser, and $\Omega_b$ and $\Omega_s$ are the beaming and subtended solid angle of the maser. For a cylindrical maser model, $\Omega_b = (d/L)^2$, and $\Omega_s = (d/D)^2$, where $d$ is the cross section diameter of the maser, $L$ is the maser length, $D$ is the distance to the maser source. Hence, we can write for $D$ = 7.2 Mpc,
\begin{equation}
R = 50 \left(\frac{S_{\nu}}{1~\rm{Jy}}\right) \left(\frac{L}{10^{16}~\rm{cm}}\right)^{-2} \rm{s^{-1}}.
\end{equation}

If the coherent velocity interval is 1 \kms, then the maximum coherent amplification length due to the line-of-sight gradient for a high-velocity maser feature at a radius of 0.2 pc is $0.2~\mathrm{pc} \times \sqrt{2~\mathrm{km~s^{-1}} / v_{\rm{rot}}}$ or  $ 3 \times 10^{16}$ cm. Hence, $R_{\rm{max}}$ is 5 s$^{-1}$ for the high-velocity masers with typical flux densities $S_{\nu} = 1~\mathrm{Jy}$. For the systemic masers, the beaming angle is known because of the clear velocity range of the systemic masers. It is 8\degr, so that $\Omega_b = 0.015 ~\mathrm{sr}$. In this case, we can write $R = 0.8~S_{\nu}~(d/10^{16} \mathrm{cm})^{-2}$. The maser spectrum is very crowded with the spacing between features of about 1\kmss~or larger. This means that $d \ga 6 \times 10^{14}$ cm, and $R$ is less than 500 for the systemic features with typical flux densities $S_{\nu} = 2~\mathrm{Jy}$.

This means that the high-velocity masers are probably not saturated. Furthermore, the linewidth of 1.01\kmss~(FWHM) of the high-velocity feature at 1269\kmss~is inconsistent with a rebroadened, highly saturated maser that is a blend of the 3 strongest hyperfine components, for the spectral line breadth at high saturation is expected to be FWHM = 1.5 \kms (NW92, \citealt{emmering94,watson03_3d}). In addition, \citet{watson00_sat} favor $R/ \Gamma \la 1$ from numerical simulations of masers in a Keplerian disk with velocity gradients. The systemic features are probably slightly saturated, and might reach full saturation.

Even if the masers are fully saturated, the upper limit on the magnetic field would be lower by a factor of maximally 4 (\citetalias[Fig. 7]{vlemmings02}; \citetalias[Fig. 7]{nedoluha92}) -- consequently, at any event, our reported magnetic field upper limits are truly upper limits. Since saturation increases the observed circular polarization, even in the event of strong saturation, the true magnetic field is smaller than the quoted upper limit.

\section{DISCUSSION}\label{discussion_sec}
\subsection{The Total Magnetic Field}\label{mhd_sec}

We obtained upper limits on the toroidal and on the radial component of the magnetic field in the accretion disk of NGC~4258 by searching for Zeeman-induced circular polarization. Based on these results we did simulations to estimate the limit on the total magnetic field. Since the toroidal and the radial component, $B_{\rm{tor}}$ and $B_{\rm{rad}}$, were probed at only slightly different radii, but different locations in the disk, we regard them as representative values for the disk and thus assume they can be used to characterize the magnetic field at a specific point, i.e., radius.

\begin{figure*}[!ht]
\epsscale{0.9}
\plotone{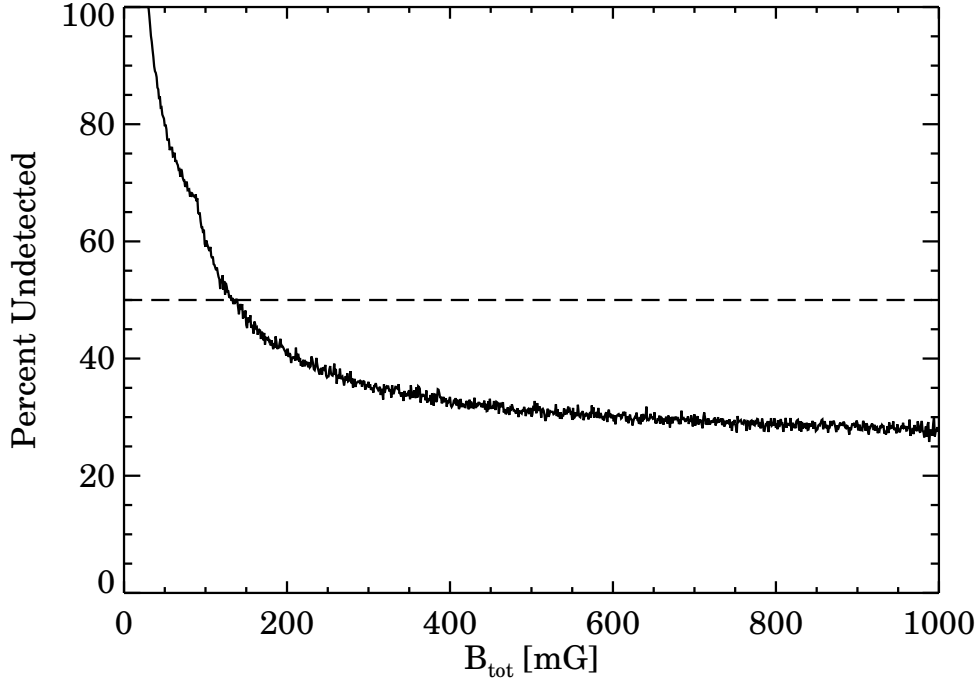}
\caption{Result of simulations of the value of the total magnetic field ($B_{\rm{tot}}$) given our measured upper limits on the two components of the $B$ field. The percentage of simulations where the total $B$ field satisfies the upper limits on the toroidal and radial component and is therefore undetected with a sensitivity comparable to our measurements is plotted as a function of $B_{\rm{tot}}$. The dashed line indicates the 50\% limit, i.e., in 50\% of the trial runs the observational limits were met (see text for details).}
\label{btotfig}
\end{figure*}

We conducted a simulation by assigning random orientations to a range of total $B$ field strengths, $B_{\rm{tot}} = 0 - 1000$ mG. For each value of $B_{\rm{tot}}$, 5000 random orientations were drawn, with uniform distributions in $B_{\rm{tor}}$, $B_{\rm{rad}}$, and $B_{\rm{pol}}$, the poloidal component.
 
For each orientation, the value of the two components of interest were compared to our observational upper limits. Figure~\ref{btotfig} presents the results of the simulations: it shows the percentage of trials in which the toroidal and radial components of the  intrinsic total $B$ field could not be detected with our observational accuracy as a function of the intrinsic total $B$ field. We find that the simulated orientations for the $B$ field would exceed measurements 50\% of the time for a B$_{\rm{tot}}$ value of 130 mG. Thus, we adopt the estimate of 130 mG as the upper limit on the total $B$ field at a distance of 0.14 pc from the central SMBH. The Alfv\'{e}n speed, $v_A = B / \sqrt{4\pi \rho}$, would be 2$-$20 \kms~for the the allowable range of densities of molecular hydrogen, $10^8 - 10^{10}$ \ccm, and for $B_{\rm{tot}} \la 130$ mG.

From simple hydrostatic equilibrium arguments where the magnetic pressure confines the masing gas, one could expect the following values for the $B$ field: From $B = \sqrt{8\pi n_{\rm{H_2}} k T_k}$, the minimum value would be $B_{\rm{min}} = 10$ mG (for the minimum temperature of $T_{k}$ = 250 K and minimum molecular number density of $n_{\rm{H_2}} = 10^8$ \ccm) and the maximum value $B_{\rm{max}} = 190$ mG (for $T_k = 1000$ K and $n_{\rm{H_2}} = 10^{10}$ \ccm). LOS magnetic field measurements in starforming regions via water maser polarization observations lie within this range (FG89, \citealt{sarma01,sarma02}), and polarization observations in envelopes of late-type starts indicate super-thermal magnetic fields of up to $\sim 0.1 - 3$ G (\citetalias{vlemmings01,vlemmings02}).

An alternate way of estimating the total magnetic field is by considering predictions from simulations of magnetic fields in accretion disk. A series of magnetohydrodynamic (MHD) simulations show that the combined presence of magnetic field and outwardly decreasing differential rotation can rapidly create MHD turbulence via the magneto-rotational instability (MRI) (\citealt{balbus91}; \citealt{stone96}; see also \citealt{balbus98_review,balbus03}). This results in a greatly enhanced effective viscosity, which readily generates outward flow of angular momentum. \citet*{hawley96_3comp} performed local three-dimensional simulations of magnetized shearing box models, with an weak ($p_{\rm{magnetic}} \ll p_{\rm{thermal}}$) and zero-mean magnetic field as part of the initial conditions. Using a variety of initial field strengths and geometries, they found that the large scale magnetic field is dominated by the toroidal component: $B_{\rm{tor}}^2 : B_{\rm{rad}}^2 : B_{\rm{pol}}^2 \approx 10:5:1$. If MHD turbulence is dominant, one might expect a magnetic field configuration as found by \citet{hawley96_3comp}. Hence, our measured upper limit on the radial component of the magnetic field then dictates an upper limit on the toroidal component of $\la \sqrt{2}~30~\rm{mG} = 43$ mG (GBT data), which is consistent with, and tighter than, the observationally determined upper limit of $B_{\rm{tor}} \la 90$ mG at a slightly larger distance (Table~\ref{zeemanfittab}). The MHD predicted total magnetic field would thus be $B_{\rm{tot}} =\sqrt{16/5}~ B_{\rm{rad}}\la$ 50 mG at a radius of 0.14 pc, where we used the limit from the GBT data on the radial component, and $B_{\rm{tot}} = \sqrt{16/10}~ B_{\rm{tor}}\la$ 110 mG at a radius of 0.2 pc, where we used the limit on the toroidal component. Since the radial limit is estimated closer to the black hole, we adopt $B_{\rm{tot}}\la$ 50 mG as the MHD predicted total magnetic field.

There might be concerns that the magnetic field is not sufficiently coupled to the gas, as the ionization fraction prevalent in the molecular masing gas is probably very low. To probe this question, \citet{menou01} conducted studies on ionization, MHD and gravitational instabilities in AGN accretion disks and found that the conditions in NGC~4258's accretion disk still satisfy the requirements for large Reynolds numbers; hence, the neutral particles are expected to be well-coupled to the magnetic field by collisions with the ions, and the MHD turbulence due to MRI in NGC 4258's accretion disk is expected to be similar to that in an ionized plasma. They also found that the thermal ionization instability is unlikely to be a source of efficient angular momentum transport and, thus, does not induce large-amplitude luminosity fluctuations in the disk. 

\subsection{Effects of the Magnetic Field}\label{effects_sec}

\citet{rees87,bottorff97_5548}, and more recently, \citet{bottorff00}, have argued that magnetic fields of strength $\sim 5-10$ G confine the clouds that populate the broad-line regions (BLRs) in AGNs at distances $R \sim 10^{17} - 10^{18}$ cm from the SMBH. \citet{ho97_halpha} report the detection of a broad H$\alpha$ component in high-resolution, high S/N total-light spectra of NGC~4258 (after decomposing the emission lines), with a FWHM of 1700 \kms. We note that the detected emission lines in polarized light (i.e., light scattered off material surrounding the central source) probably emanate from the narrow-line region, not the BLR \citep{wilkes95,barth99}. From the virial estimate, $R_{\rm{BLR}}$, the position of the BLR, is approximately $M_{\rm{BH}}G/v^2$, with $v = f v_{\rm{FWHM}}$, where the factor $f$ depends on the details of the geometry, kinematics, and orientation of the BLR and is $\sqrt{3}/2$ for an isotropic distribution with random inclinations \citep{netzer90}. Thus, $R_{\rm{BLR}} \sim 2.4\times 10^{17}$ cm or 0.08 pc. This is smaller by a factor of $2-3$ than the distance regime the magnetic field upper limits reported here are probing. In order to confine the BLRs at that distance, the $B$ field strength would have to scale as $r^{-4}$ to reach the required high field strengths --- since this scaling is unphysical, we conclude that magnetic pressure is unlikely to be the dominant force to confine a BLR of radius 0.08 pc in NGC~4258 (see, however, \citealt{laor03} for arguments on why NGC~4258 might not harbor BLRs). However, since this system is an obscured AGN, the width of the broad H$\alpha$ component could be larger, meaning the BLRs could be located closer to the SMBH and thus high magnetic fields could be reached at those distances. We note that our upper limits reflect $B$ fields in very dense masing gas and that the postulated $B$ fields responsible for confining the BLR clouds would be presumably lower as they pervade a more rarefied intercloud medium (be it relativistic or accretion-driven wind or accretion flow; \citealt{rees87}).

 Large scale poloidal fields may play an important role in collimating AGN jets (see, e.g. \citealt*{begelman84}). VLBI observations of M87, which contains one of nearest large-scale jets, show that the jet is well-collimated at $30-100$ Schwarzschild radii, and are interpreted as suggesting an accretion disk threaded by a strong poloidal magnetic field as the launching mechanism \citep*{junor99_m87}. The poloidal $B$ field component in NGC 4258's accretion disk, which is observationally inaccessible, can be estimated to be $B_{\rm{pol}} \sim B_{\rm{rad}} / \sqrt{5}\la $ 14~mG (GBT data) at a radius 0.14 pc (or 37,000 $R_S$) from MHD model constraints. Since for a steady-state thin accretion disk, which is in thermal equilibrium, the surface density goes as $r^{-3/4}$ \citep*{fkr}, the upper limit on $B_{\rm{pol}}$ at $r$ = 0.14 pc translates to $B_{\rm{pol}} \la~$7 G at 10 $R_S$, for an $\alpha$-thin disk that extends inwards to $\sim$10 $R_S$. This is much less than the magnetic fields of strength $10^2-10^4$ G required for collimating radio jets \citep{blandford82}.

\subsection{Mass Accretion Rate}\label{mdot_sec}

Relating the thermal pressure at the midplane of the accretion disk to the magnetic pressure, the mass accretion rate can be estimated from the magnetic field strength. The thermal pressure on the midline is given by (e.g., \citealt{neufeld95,herrnstein98_pol}):
\begin{equation}
P_{\rm{th}} = \rho c_{s}^2 = \frac{1}{3\pi \sqrt{2\pi}}\frac{G M \dot{M}}{ c_s r^3 \alpha},
\label{pth_eq}
\end{equation}
where $\alpha$ is the Shakura-Sunyaev viscosity parameter, and $M$ the mass of the black hole. 

Setting equation~(\ref{pth_eq}) equal to $\beta B^2 / 8 \pi$, where the parameter $\beta$ is the ratio between the gas and the magnetic pressure $\beta = 8 \pi \rho c_s^2 /B^2 $ and is 1 for equipartition (note that $\beta$ is sometimes defined as $\beta = P_{\rm{tot}} / P_{\rm{mag}}$), we obtain

\begin{equation}
\dot{M} = \frac{3\sqrt{2\pi}}{8}~\frac{\alpha \beta  B^2 c_s r^3 }{G M}~.
\label{mdot_eq}
\end{equation}

\noindent If magnetic fields provide the kinematic viscosity, the parameters $\alpha$ and $\beta$ are likely to be related according to the MHD-simulations prescription $\alpha = C/(1+\beta)$, where $C$ is usually set to $0.5-0.6$ (\citealt{narayan98}, note their slightly different definitions of $\beta$ and of the magnetic pressure, which are accounted for in the above equation). In that context, an equipartition magnetic field corresponds to $\alpha \sim 0.3$. Local three-dimensional MHD calculations suggest sub-thermal magnetic fields ($\beta \approx 30 - 100,$ $<$$\beta$$>$~$\approx$60, \citealt{hawley96_3comp}).  Recent global three-dimensional MHD simulations indicate a large magnetic field contribution, with $\beta \approx 1 - 5$ \citep{hawley00}, because the length-scales, where MRI can be effective, are larger in the global simulations than in local ones. However, they are conducted at radii very close to the SMBH ($\sim 1-10~R_S$), owing to computational constraints, deal with torii at those distances ($H / R \sim 1$), and thus do not apply to our situation of a thin disk. 

Adopting a temperature of 400 K in the disk midplane, as suggested by analyses of observed water maser line profiles, equipartition between magnetic and thermal pressure, and a $B$ field upper limit of 130 mG from the polarization analysis of the various maser lines and statistical considerations, we derive at the radius of $r$ = 0.14 pc a mass accretion rate of 

\begin{eqnarray}
 \dot{M}& \la 10^{-3.7}~
\frac{0.6~(\beta / 1)}{1+(\beta / 1)}
\left(\frac{B_{\rm{tot}}}{\rm{130~mG}}\right)^2 
\left(\frac{T_k}{\rm{400~K}}\right)^{1/2} & \nonumber \\
&\left(\frac{r}{0.14 ~\rm{pc}}\right)^3
\left(\frac{M_{\rm{BH}}}{3.9 \times 10^7~\rm{M}_{\sun}}\right)^{-1}~
\rm{M}_{\sun} yr^{-1}.& 
\label{mdotnum_eq}
\end{eqnarray}


If we adopt the magnetic field arrangement as predicted by \citet{hawley96_3comp} with subthermal fields, such that B$_{\rm{tot}}^2 = 16 B_{\rm{rad}}^2 / 5$, and $\beta \approx$ 60 (and thus $\alpha = 0.01$), we obtain $\dot{M} \la 10^{-4.2}~\rm{M_{\sun}yr^{-1}}$ at a radius of 0.14 pc. We stress that this analysis is based on an upper limit on the $average$ magnetic field across the masing region of $\sim 7 \times 10^{15}$ cm \citep{moran95}; if the magnetic fields are significantly tangled, then the $B$ field limit is underestimated by a factor of a few (\citealt{watson01}, see also \S~\ref{saturation_sec}). 

The estimate in equation~(\ref{mdotnum_eq}) reflects an improvement by one order of magnitude from the previous estimate of the mass accretion rate in NGC~4258's disk (scaled to the presently estimated and measured values of $M_{\rm{BH}}$, $T_k$ and $D$) using the same methods \citep{herrnstein98_pol}. Our result is consistent with the mass accretion rate of 10$^{-4} \alpha~\rm{M_{\sun} yr^{-1}} $ currently estimated by J.R. Herrnstein et al. (2004, in preparation), which is based on considerations of the transition between molecular and atomic gas using the X-ray irradiation model of \citet{neufeld95} and incorporates new VLBI observations of the masers.

The disk mass can be estimated by considering the deviations of the maser velocities from pure Keplerian motion (3\kmss), which indicates that the mass pertained in the disk is less than about $10^6 M_{\sun}$ \citep{moran99}. The associated average time-scale for accreting the whole disk is $t = M_{\rm{disk}}/\dot{M} \sim 10^{10}$ years, while the viscous time scale (i.e., the time scale for matter to diffuse through the disk from the masing region under the effects of viscous torques) is $t_{\rm{visc}} \sim R/v_R$ with $v_R <0.006\alpha$ \kmss~from the observations of the systemic features \citep[since $v_R = \alpha v_{\phi}(H/R)$]{moran99}, such that $t_{\rm{visc}} \sim 4 \times 10^7$ years (for $\alpha = 0.3$) and $t_{\rm{visc}} \sim 1 \times 10^9$ years (for $\alpha = 0.01$). For comparison, the Eddington accretion rate for a central black hole mass of $M_{\rm{BH}} = 3.9 \times 10^7 M_{\sun}$ is $\dot{M}_{Edd} =  0.9 ~ \rm{M_{\sun}yr^{-1}}$ (e.g. \citealt{peterson97}, assuming a Shakura-Sunyaev disk), where the radiative efficiency parameter has the fiducial value of 0.1. Thus, our derived mass accretion rate is $\dot{m} \equiv \dot{M} / \dot{M}_{Edd}  \la 10^{-3.7}$ for $\beta = 1$ and $\dot{m} \la 10^{-4.1}$ for $\beta = 60$.

The above discussion relies on the assumption that the masers occupy the midplane of the accretion disk, for equilibrium considerations to be valid. There is no direct observational evidence for this claim, and in some models the masers inhabit a surface layer above a much colder and more massive accretion disk (\citealt*{desch98}) or are uplifted by a disk-driven hydromagnetic wind \citep{kartje99} (\S~\ref{masermodels_sec} gives a discussion of these alternate models). Nevertheless, the simple model of a warped Keplerian model and its characteristics fit remarkably well the observational data (\citealt*{herrnstein96}, \citealt{herrnstein97_phd}, \citealt{herrnstein97_jet}, \citealt{herrnstein99}, \citealt{humphreys04}). Any potential surface layer has to be very thin, $H \la 15~\mu$arcsec or 0.0005 pc (from the spatial distribution of the masers in the systematic group, \citealt{moran95,herrnstein05}). For the systemic features, there is no direct way to measure any potential displacement of the masers from the midline. We can attempt to provide a rough estimate by considering the radio continuum emission at 22 GHz from northern and southern lobes of the nuclear jet. If we take the midpoint connecting the two lobes as shown in \citet{herrnstein97_jet} as the location of the SMBH and compare it to the position of the dynamical center of mass as predicted by the maser kinematics, we estimate a difference of less than about 150 $\mu$arcsec. If we take this limit as the offset between the maser layer and the midline (150 $\mu$arcsec), i.e., 10 scale heights, we conclude that the mass accretion rate upper limit (eq.[\ref{mdotnum_eq}]) is larger by 15 orders of magnitude, for a vertical pressure profile of $P (z) \propto (1- Erf(z/\sqrt{2}H))^{1/4}~\mathrm{exp} (-z^2/2H^2 ) $, where $Erf(x)$ is the standard error function (see, e.g. \citealt{goodman03}). However, this estimate is most likely flawed, as the emission of the southern jet is probably attenuated by a layer of ionized gas in the disk via free-free or synchrotron absorption \citep{herrnstein97_jet} and thus, the true centroid of the southern lobe may not be revealed in the 22 GHz maps. High-resolution submillimeter maps in the future era of ALMA could alleviate the opacity problem and might answer the question fully. If the masers indeed arise on a surface layer and are not located on the midline, one would expect a discontinuity at the position of the black hole for an anti-symmetrically warped disk. The emission sites of the redshifted high-velocity features would be expected to lie on the top on one side, and the maser-skin of the blueshifted high-velocity features on the bottom on the other side. 

\subsection{Comparisons to Maser Model Accretion Rates}\label{masermodels_sec}

More than 50 AGNs with megamaser emission are currently known (L. Greenhill, private communication, 2004, see also \citealt{braatz03,greenhill03}), and several show clear signs of disk structure. NGC~4258 is one of the best-studied cases and provides many observational details, and numerous mechanisms have been proposed to explain the production of maser emission in its circumnuclear environment.

\citet*{neufeld94} and specifically \citet{neufeld95} propose irradiation of the warped circumnuclear accretion disk by X-rays from the central AGN as the mechanism to produce water masers. \citet{neufeld95} show that the outer edge of the maser distribution at $r \sim$ 0.28 pc can be explained in terms of a phase transition from molecular to purely atomic gas, provided that the accretion rate is $\dot{M}\approx10^{-4.1}~\alpha~\rm{M_{\sun}yr^{-1}}$ at a radius of 0.28 pc (see also \citealt{maloney02}). Their mass accretion rate is in very good agreement with our upper limit. 

The inner radial cut-off of the maser annulus is postulated to be due the absence of the disk warp at the smaller radii and consequently, the absence of X-ray heating. Long-term monitoring observations of NGC~1068, an archetypical Seyfert II megamaser galaxy, show that the flux density of the nuclear masers are fluctuating coherently on timescales of months to years \citep{gallimore01}. These variations occur faster than the dynamical timescale would allow and can be invoked as response in reverberation to the central engine as the pump source, thus fitting the scenario of central X-ray heating (however, it appears that the X-rays experience heavy absorption and can only be detected in scattered light; \citealt{pier94}). Similar monitoring campaigns of simultaneous X-ray and maser emission observations are underway for NGC~4258 (L. Greenhill 2004, private communication).

As an alternative to the warped disk interpretation of megamasers, \citet*{kartje99} propose a scenario where the masing gas is confined and uplifted from a flat accretion disk by a powerful, dusty, centrifugally driven MHD wind, which exposes the gas to X-ray irradiation and excites maser emission. This model proposes that the apparent warp seen in the maser disk of NGC~4258 is due to a non-coplanar distribution of masing clouds and stems from their argument that the maser regions are clumpy, after considering radiative and kinematic aspects and the time variability of the observed maser flux. The nearly perfect Keplerian shape of the rotation curve as traced by the maser features is explained qualitatively in terms of observational selection effects, but without fitting the wind-model to the observed spatial distribution of maser spots and their line-of-sight velocities. It is hard for this model to explain the anti-symmetric shape of the warp in the almost perfect Keplerian rotation curve of NGC~4258. Also, it is unclear how other observational trends can be explained with this model (e.g., the relative strengths of the two groups of high-velocity features, the apparent small height of the disk as traced by the masers). Their estimated line-of-sight $B$ field of $\sim$20 mG in the homogeneous outflow component at $r\sim$0.2 pc is consistent with our upper limit of 90 mG at a similar position; furthermore, the $B$ field in the masing cloud is expected to be smaller than the estimate in the outflow-wind, since the clouds are assumed to be diamagnetic (i.e., not permeated by an ambient magnetic field) in their model. Their derived mass accretion rate at $r = 0.2$ pc is $\dot{M}\approx 10^{-1.1}~\rm{M_{\sun}yr^{-1}}$ (their eq. [29]), assuming that 10\% of the mass is carried away in the wind, and inconsistent with our value of $\dot{M} \la 10^{-3.7}$. Of course, a detailed comparison is not warranted here, since our accretion rate computation assumes that the masers lie in the accretion disk, the very opposite picture of the model by \citet{kartje99}.

As an alternative to the X-ray heating scenario, \citet{maoz98} envision spiral density waves sweeping through a self-gravitating accretion disk and triggering shock waves that heat the gas. This model is motivated in part to explain the consistently larger flux densities of the red-shifted over the blue-shifted high-velocity lines and the periodic spacing of the high-velocity features in radius and velocity. They predict a mass accretion rate of $\dot{M} \approx 10^{-2.2}~\rm{M_{\sun}yr^{-1}}$ at the location of the masers ($\sim$0.2 pc) from energy dissipation arguments in the spiral density shocks. This is larger by almost two orders of magnitude than our present upper limit for the accretion rate at the same radius and thus, is not consistent with our results. Here, a direct comparison is warranted as the same disk geometry is assumed and prediction about the mass accretion rate concerns the same location in the disk. \citet{bragg00} find no observational evidence for the shock-model-predicted pattern in the accelerations of the high-velocity features. In addition, \citet{herrnstein96} show that the persistent asymmetry in the spectrum can be well-explained by a warped and X-ray irradiated disk. 

\subsection{Comparisons to Accretion Models}

Provided accretion has been constant over the time span needed for material to reach the galactic nucleus from the maser disk, the estimate reported in this work can be used to probe the nature of the accretion responsible for the energy output of the central AGN. The viscous timescale is t$_{\rm{visc}} \sim 4 \times 10^7$ years (for $\alpha = 0.3$) or t$_{\rm{visc}} \sim 10^9$ years (for $\alpha$ = 0.01; see \S~\ref{mdot_sec}) and it is not certain whether the disk has been in steady state over the whole time range. For a jet whose evolution is closely linked to the evolution of the accretion disk, the dynamical life time of the jet and its relatively undisturbed morphology gives an estimate of $t_J \sim 10^6$ years \citep{cecil95_jettime} as a time frame of steady accretion.
\begin{figure*}[!ht]
\epsscale{0.9}
\plotone{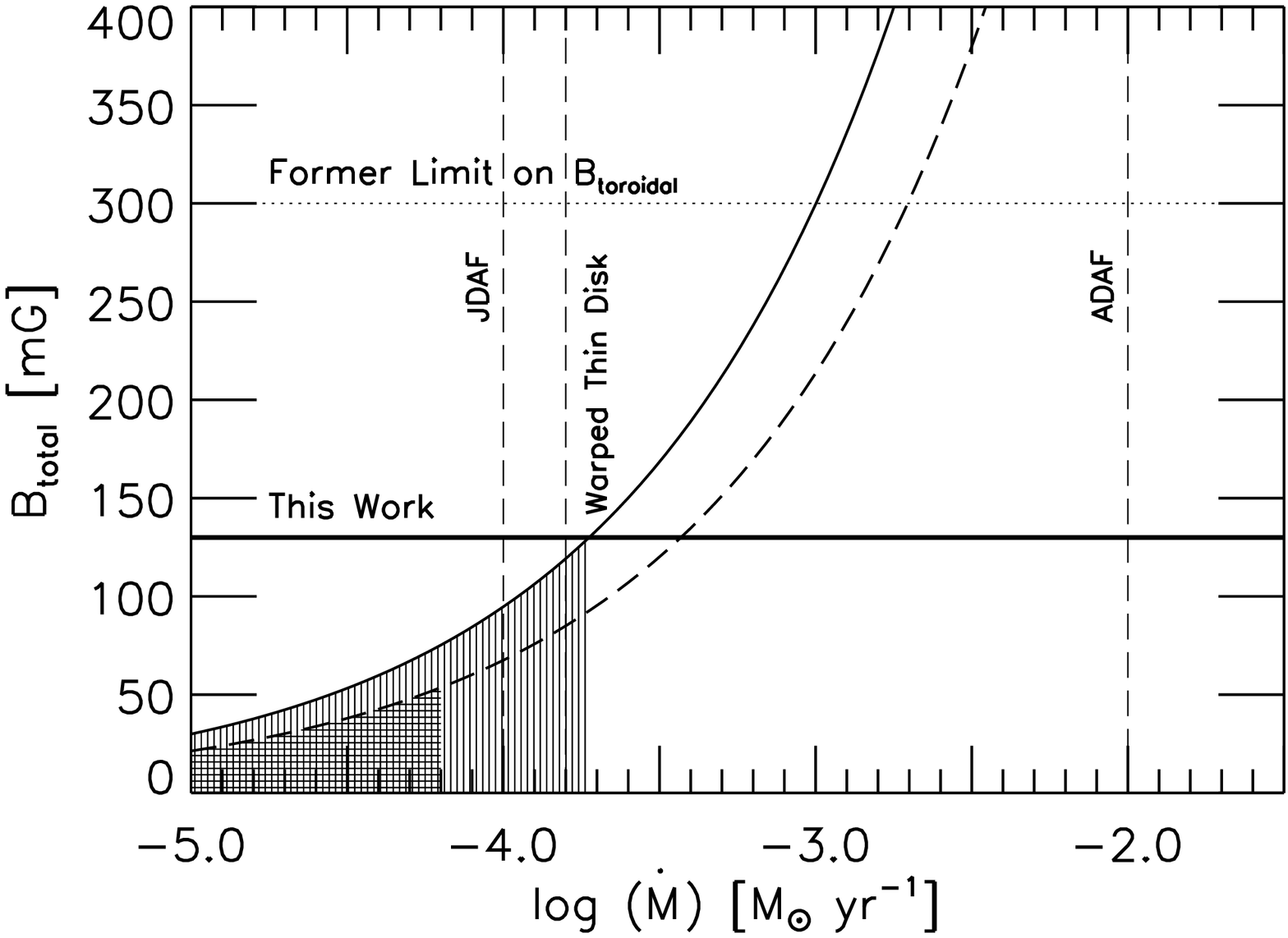}
\caption{Expected strength of the total $B$ field as a function of mass accretion rate (eq.~[\ref{mdotnum_eq}]) at a radius of 0.14 pc for T$_K$ = 400 K, and for $\beta = 1$ (solid curved line) and $\beta = 60$ (dashed curved line). The accretion rates predicted by the jet-disk model of \protect\citet{falcke99} (JDAF), the thin disk and the ADAF models of \protect\citet{gammie99} are shown (dashed vertical lines), as well as the upper limit by \protect\citet{herrnstein98_pol} and that from the present experiment (horizontal lines). The \it{vertically}\rm~shaded region shows the constraint on the accretion rate assuming equipartition magnetic fields (i.e., $B_{\rm{tot}}\la 130$ mG, $\beta = $1 and, thus $\alpha = 0.3$) and the \it{cross hatch}\rm~shaded region depicts the constraint having incorporated the MHD predictions by \protect\citet{hawley96_3comp} (i.e., B$_{\rm{tot}} = \sqrt{16/5}~ B_{\rm{rad}}\la$ 50 mG, $\beta = $60 and, thus $\alpha = 0.01$, see \S~\ref{mhd_sec} for details).}
\label{bvsmdotfig}
\end{figure*}
The bolometric luminosity can be estimated from the X-ray luminosity, since $L_{X}\sim 0.1L_{\rm{bol}}$. A number of X-ray observations have been conducted over the past several years and yield values for the absorbing column density of $N_H \sim (6 - 15) \times 10^{22}$ cm$^{-2}$ and for the $2-10$ keV absorption corrected luminosity of $L_X = (4 - 13)\times 10^{40}$ erg~s$^{-1}$ (for $D$ = 7.2 Mpc) (see \citealt[Table 2]{fruscione05} for a compilation and references therein). Thus, $L_{\rm{bol}} \sim 10^{42 \pm 1}$ erg~s$^{-1} = 10^{-3.6 \pm 1}L_{\rm{Edd}}$, where $L_{\rm{Edd}} = 1.3 \times 10^{38} (M_{\rm{BH}}/M_{\sun}) $\ergs~is the Eddington luminosity for this system (but see \citealt{woo02,laor03} and for alternate estimates of the bolometric luminosity).

The highly sub-Eddington nature of the central source in NGC~4258 could be explained by an $\alpha$ disk that is geometrically thin, optically thick and reaches to the central black hole and whose low mass accretion rate leads to the sub-Eddington luminosity of the central engine. If the observed nuclear jet as seen in the radio is powered by the accretion flow (and not by extraction of the spin energy of the SMBH), the minimum mass accretion rate has to be $\dot{M} \sim 1.2 \times 10^{-3}~\epsilon_{-1}^{-1}~ \beta_e^{0.4}~(\Gamma / 10)^{3.4} ~\rm{M_{\sun}yr^{-1}} $ \citep*{gammie99}, where $\beta_e$ is the ratio between the electron and the magnetic energy density (assumed to be $\ga$ 1 for jets) and $\Gamma$ the Lorentz factor of the relativistic electrons, which is $\ga 5$ (P. Kondratko, 2004, private communication, \citealt{herrnstein97_jet}). Thus, the accretion rate estimated in this paper would be sufficient to power the nuclear yet. We note that the standard thin disk model alone cannot generate the large luminosity seen in X-ray portion of the SED, unless an X-ray emitting, hot, tenuous corona is present as, e.g. suggested by \citet{haardt91} for Seyfert galaxies.


Alternatively, a number of authors have proposed that NGC~4258 is powered by an Advection-Dominated Accretion Flow (ADAF)\citep{lasota96,gammie99,chary00}. The ADAF model (e.g., \citealt{narayan94}; see \citealt*{narayan98} and references therein for review) requires a geometrically thick, but optically thin disk, with a two-temperature plasma of hot ions ($\sim10^{11}$ K) and cooler electrons ($\sim 10^9$ K). The high temperatures of the ions are maintained due to the inefficiency of collisional energy transfer (Coulomb interactions) at the expected high temperatures and low densities, i.e., accretion rates. Since the ADAF cannot cool radiatively, it ``puffs'' up ($H / R \sim 1$), and the majority of the viscously dissipated energy gets transported by the ions into the black hole. The ADAF system emits at sub-Eddington luminosities due its low radiative efficiency of the energy-carrying ions and has been put forth in explaining low-luminosity AGNs (such as NGC~4258), which might occupy a significant fraction of the local galaxy population \citep{ho95}. \citet{gammie99} fit an ADAF model to the spectral energy distribution (SED) of NGC~4258 while incorporating the stringent 22 GHz upper limit by \citet{herrnstein98_cont} and the IR data by \citet{chary97}, which constrain the transition radius between ADAF and thin disk to lie at $5-50 R_S = (10-100) GM/c^2$. Their derived mass accretion rates for slightly different ADAF models cluster around $\dot{M} = 10^{-2}~\rm{M_{\sun}yr^{-1}}$. This ADAF accretion rate is not consistent with our reported accretion rate.

Figure~\ref{bvsmdotfig} shows a comparison of the different $\dot{M}$ values including our upper limit on the total $B$ field from our measurements in conjunction with statistical considerations and from MHD predictions.
Detection of a relativistically broadened K$\alpha$ emission line at 6.4 keV would conclusively demonstrate the presence of cold material close to the last stable orbit in NGC~4258 and rule out the standard ADAF model. Detection claims have been made of a $narrow$ iron component \citep*{reynolds00} that is apparently time-variable \citep{pietsch02,fruscione05}, but there has been no detection of a broad emission component.


Nonetheless, since both ADAF and thin disk models predict a spectral behavior of $F_{\nu} \propto \nu^{1/3}$ over the IR portion of the SED (from the emission of the thin disk), they are in conflict with the most recent high-resolution mid-IR data obtained by \citet{chary00} that indicate a non-inverted and a much steeper power-law ($F_{\nu} \propto \nu^{-1.4}$). A jet-dominated-accretion-flow, which has been proposed as an alternative to both the ADAF and the thin disk model \citep{falcke99,yuan02}, is able to accommodate the large IR flux and its spectral behavior. The compact jet that is evoked to give rise to the whole extent of the SED would be fueled by an outer ADAF and a further-out lying thin accretion disk. The mass accretion rate that is needed to power the jet appears to be low: \citet{falcke99} estimated an accretion rate of the order $10^{-4}~\rm{M_{\sun}yr^{-1}}$ and \citet{yuan02} showed that jet mass loss rates of $\sim (0.7 - 2)\times 10^{-4}~\rm{M_{\sun}yr^{-1}}$ generate viable JDAF spectra in agreement with data over a broad range of energies. The exact relationship between jet loss and disk accretion rate is uncertain, nevertheless, \citet{yuan02} show that low accretion rates ($\dot{M}\sim 10^{-(4-5)}~\rm{M_{\sun}yr^{-1}}$) are required in order to have negligible emission from the disk and the ADAF. These values are are consistent with our estimated accretion rate (Fig.~\ref{bvsmdotfig}).

There exist a number of other accretion scenarios that belong to the larger class of radiatively inefficient accretion flows (which include ADAFs and JDAFs) and that could explain the accretion processes operating in the central engines of low-luminosity AGNs: Convection Dominated Accretion Flows (CDAFs; \citealt*{narayan00,quataert00}) and Adiabatic Inflow Outflow Solutions (ADIOS; \citealt{blandford99}). As these models have not been fit to the spectrum of NGC~4258 in detail, they will not be discussed here at length. Qualitatively, the CDAF model is expected to be mass-starved, since the inward transport of angular momentum by convection chokes accretion. The ADIOS model stipulates a powerful wind, which carries away energy and angular momentum from the accreted gas; thus it would require a large accretion rate at large distances from the SMBH to compensate for the mass loss occurring further in, leading to a low mass accretion near the event horizon.

\section{SUMMARY AND CONCLUSIONS}\label{conclusions_sec}

In this paper, we have presented total power and polarization spectra of the nuclear maser in NGC~4258 taken with the VLA and the GBT. These spectra constitute the most sensitive data with the most complete simultaneous velocity coverage of maser emission in this system. Based on the analysis of the polarimetric observations we conclude the following:

1) There is no circular polarization indicative of Zeeman-induced splitting of the maser lines of water vapor, a non-paramagnetic molecule, at the level of 0.1 $-$ 0.3\%. This leads to a 1 $\sigma$ upper limit on the toroidal component of the magnetic field of 90 mG (at 0.25 pc, at the position of the redshifted high-velocity features) and on the radial component of 30 mG (at 0.14 pc, at the position of the systemic features) in the maser-delineated accretion disk of NGC~4258. No linear polarization at a level of $\sim$0.1\% is detected. If the masers are saturated, the $B$ field upper limits may be lower than reported here. 

2) The magnetic field upper limits are less than the several-Gauss-fields required to magnetically confine the broad line regions clouds, whose emission is seen in high-resolution, high signal-to-noise optical data in NGC~4258 \citep{ho97_halpha}, provided that the clouds are close to the position of the maser features (as suggested by the virial estimate).

3) Assuming equipartition between magnetic and thermal energy in the accretion disk (and thus $\alpha$ = 0.3) and that the masers are situated on the midline, we derive a mass accretion rate of $\la10^{-3.7} M_{\sun}$yr$^{-1}$ for a total magnetic field of less than 130 mG. Alternatively, adopting the $B$ field configuration as predicted by MHD simulations of \citet{hawley96_3comp}, we obtain $ \dot{M}\la10^{-4.2} M_{\sun}$yr$^{-1}$. Both upper limits on the accretion rate are in agreement with estimations from the X-ray irradiation maser models by \citet{neufeld95}, but do not agree with the high $\dot{M}$ from the spiral density model by \citet{maoz98} predicted at the same radius.

4) Provided the accretion disk has been in steady state and the masers reside on the midline, the low accretion rate reported here favors mass-starved models as candidates for explaining the extremely low-luminosity nature of NGC~4258. These models include the thin-disk model and the jet-disk model.

Detection of the magnetic field in the accretion disk of NGC~4258 could be achieved with more sensitive data. It would be especially important to obtain detections in the numerous high-velocity features in order to map the magnetic field structure as a function of radius in an AGN accretion disk. MHD theories could be tested, as they predict random, not ordered, magnetic fields at different radii. Detection of the radial component of the magnetic field via the systemic features might be hampered by saturation effects. 

\acknowledgements
We would like to thank Greg Ball, Charles Gammie, Ramesh Narayan, Mark Reid, Wouter Vlemmings, and Bill Watson for fruitful discussions and illuminating correspondence, and Jim Braatz and Ron Maddalena for their support at the GBT. This work was supported in part by NRAO grant GSSP02-0006 and has made use of NASA's Astrophysics Data System Bibliographic Service. 
 
\clearpage
\newpage

\appendix

\section{Comparison to Thermal Zeeman Effect Terminology in Radio Astronomy}\label{thermalz_sec}

Another instance of fractional Zeeman splitting can be found in $thermal$ OH or \ion{H}{1} emission or absorption observations. Even though the Zeeman splitting coefficients are $\sim10^3$ times larger for those paramagnetic molecules than for water, the regions in which the emission originate are of low density, such that the $B$ fields are in the range of $\mu$G, hence producing small Zeeman splittings. Numerous thermal Zeeman effect radio experiments thus fit the following expression to the Stokes $V$ spectrum (see \citealt{troland82,sault90} for good reviews):
\begin{equation}
V(v) = c_1~\frac{dI(v)}{dv} + c_2 I(v) .
\end{equation}
Here, $c_2$ denotes the instrumental polarization term due to small calibration errors between LCP and RCP. The factor $c_1$ is a measure of the Zeeman splitting and corresponds to $c_1 = A_{\rm{F-F'}} B \rm{cos}\theta / \sqrt{2}$. This notation is equivalent to the terminology introduced by \citetalias{fiebig89} (eq.~[\ref{zeeman_eq}]), as shown in the following paragraph.

We assume that the Stokes $I$ profile can be expressed as a simple Gaussian line: Stokes~$I$~=~$I_{\rm{o}}$ e$^{-v^2 / 2\sigma^2}$ (for simplicity, the center velocity is taken to be at 0 \kms). For the small Zeeman splitting case, the Stokes $V$ profile is proportional to the Stokes $I$ profile, or more specifically \citep{goodmanphd,crutcher93,herrnstein97_phd},
\begin{eqnarray}
 	V (v)   & \simeq & c_1	\frac{dI(v)}{dv}  \nonumber \\
		& = & I_{\rm{o}}~\frac{c_1}{\sigma^2} ~v ~e^{-\frac{v^2}{2\sigma^2}},
\end{eqnarray}

Thus, the $S$ curve is an antisymmetric curve and has extrema at $\pm \sigma$ (not at $\pm \sigma / \sqrt{2}$ as claimed by \citealt{vlemmings01,vlemmings02}) with values of $\pm (I_{\rm{o}} c_1/\sigma)~e^{-1/2} $. 

Hence, 
\begin{eqnarray}
\frac{V_{\rm{max}}}{I_{\rm{max}}} & = & \frac{V_{\rm{max}}}{I_{\rm{o}}} = \frac{c_1}{\sigma}~e^{-1/2} \nonumber \\
			& = &\frac{A_{\rm{F-F'}}}{\sqrt{2}}~B \rm{cos}\theta~\frac{e^{-1/2}}{\sigma}.
\end{eqnarray}
Since the FWHM of a Gaussian profile is defined as FWHM $= 2\sigma\sqrt{2~\rm{ln}2} = \Delta v_L $, we can substitute $\sigma$ and simplify:

\begin{eqnarray}
\frac{V_{\rm{max}}}{I_{\rm{max}}} & =& \frac{A_{\rm{F-F'}}}{\sqrt{2}}~B \rm{cos}\theta~\frac{2\sqrt{2~\rm{ln}2}}{\Delta v_L}~e^{-1/2} \nonumber \\
			&=& \frac{A_{\rm{F-F'}} B\rm{cos}\theta}{\Delta v_L}.
\end{eqnarray}

Thus, we have arrived at equation~(\ref{zeeman_eq}).

\section{Cross Correlation Method}\label{crosscor_subsec}

When the magnetic field is constant across the spectrum we can estimate the Zeeman splitting across an ensemble of lines by cross correlating the LCP and RCP spectra without forming the Stokes $V$ spectrum. Here we show that the sensitivity of this method is equivalent to that of the $S$ Curve method.

Determining the net velocity shift between the LCP and the RCP spectra can be achieved by simply finding the maximum in the cross correlation function. Since the Zeeman splitting is very small for the expected sub-Gauss $B$ fields, the maximum of the lag function will lie between lags of 0 and +1 channels or lags of 0 and $-$1 channels. Assuming a parabolic shape for the cross correlation function at maximum, we can write the following 3 equations: 
\begin{eqnarray}
CC_{+} &= a &(\Delta-q)^2 + b \nonumber \\
CC_{0} &= a &(q)^2 + b  \nonumber\\
CC_{-} &= a &(-\Delta-q)^2 + b,
\label{cc_3eq}
\end{eqnarray}
where $\Delta$ denotes the channel spacing (in\kmss) and $q$ the velocity at which the cross correlation function peaks. Here, $CC_+$ is the value of the cross correlation function at a lag channel of +1, i.e.,  
\begin{equation}
CC_{+}= \sum_{i=1}^{N} \rm{x}_i ~ \rm{y}_{i+1}, 
\label{cc_eq}
\end{equation}
with $\rm{x}_i$ denoting the value of the LCP spectrum in the i$^{th}$ channel, $\rm{y}_i$ denoting the value of the RCP spectrum in the i$^{th}$ channel and $N$ the number of channels. Analogously, $CC_0$ corresponds to the value of the cross correlation function at a lag channel of 0, and  $CC_-$ at a lag channel of $-$1. Solving the 3 equations with 3 unknowns, the net velocity shift $q$ can be obtained with
\begin{equation}
	q = \frac{\Delta}{2}~\frac{CC_{+} - CC_{-}}{CC_{-} + CC_{+} - 2~CC_{0}}.
\label{q_eq}
\end{equation}

We can obtain the conversion coefficient from velocity shift to magnetic field from the synthetic spectra (similarly to the $S$-curve method calibration by \citetalias{fiebig89}), with
\begin{equation} 
B_{||} = \frac{q}{\sqrt{2}A_{\rm{F-F'}}}, 
\label{q2B_eq}
\end{equation}
where $A_{\rm{F-F'}}$ = 0.013\kmsG for the $F = 7-6$ transition. However, we adopt the value derived from radiative transfer calculation by \citetalias{nedoluha92} of $A_{\rm{F-F'}} = 0.020$\kmsG, appropriate for the merging of the 3 strongest hyperfine components for masers.

\subsection{The Uncertainty in $B_{||}$ from the Cross Correlation Method} 

In equation~(\ref{cc_eq}), each channel value in the spectrum is a measured variable. We derive the uncertainty in the estimate of $q$ by using the standard error propagation equation:

\begin{equation}
\sigma_{q}^2 = \sum_{j=1}^{N}~(\frac{\partial q}{\partial \rm{x}_j})^2~\sigma_{x}^2 + (\frac{\partial q}{\partial \rm{y}_j})^2~\sigma_{y}^2 
+ 2 \frac{\partial q}{\partial \rm{x}_j}\frac{\partial q}{\partial \rm{y}_j}~\sigma_{xy}.
\end{equation}

Since the LCP and RCP receivers are independent, the noise in the LCP and RCP spectra are not correlated, and the covariance term $\sigma_{xy}$ can be set to zero. 

Now, define
\begin{eqnarray}
\rm{f} =& CC_{+} - CC_{-} = \sum_{i=1}^{N} (\rm{x}_i~\rm{y}_{i+1} - 
\rm{x}_i~\rm{y}_{i-1})& \\
\rm{g} =& CC_{-} + CC_{+} - 2~CC_{o}= \sum_{i=1}^{N} (\rm{x}_i~\rm{y}_{i-1} + 
\rm{x}_i~\rm{y}_{i+1} - 2~\rm{x}_i~\rm{y}_{i}).
\end{eqnarray}

Thus,
\begin{eqnarray}
\frac{\partial q}{\partial\rm \rm{x}_j} &=& 
\frac{\Delta}{2}~
\frac{\frac{\partial \rm{f}}{\partial \rm{x}_j} \rm{g} - \frac{\partial \rm{g}}{\partial \rm{x}_j} \rm{f}}
{\rm{g}^2}
=\frac{\Delta}{2}~\frac{(\rm{y}_{j+1} - \rm{y}_{j-1})~\rm{g} - (\rm{y}_{j-1} + \rm{y}_{j+1} - 2~\rm{y}_i)~f}{\rm{g}^2} \\
\frac{\partial q}{\partial\rm \rm{y}_j} &=&
\frac{\Delta}{2}~\frac{(\rm{x}_{j-1} - \rm{x}_{j+1})~\rm{g} - (\rm{x}_{j-1} + \rm{x}_{j+1} - 2~\rm{x}_i)~f}{\rm{g}^2}.
\end{eqnarray}

Finally, 
\begin{eqnarray}
\sigma_{q}^2 &=& \sum_{j=1}^{N}~(\frac{\partial q}{\partial \rm{x}_j})^2~\sigma_{x}^2 + (\frac{\partial q}{\partial \rm{y}_j})^2~\sigma_{y}^2 \\
	& =& \frac{\Delta^2}{4g^4}~\sum_{j=1}^{N}~[(\rm{y}_{j+1} - \rm{y}_{j-1})~\rm{g} - (\rm{y}_{j-1} + \rm{y}_{j+1} - 2~\rm{y}_i)~f~]^2~\sigma_x^2~+ \\
	& & ~~~~~~~~~~~~~[(\rm{x}_{j-1} - \rm{x}_{j+1})~\rm{g} - (\rm{x}_{j-1} + \rm{x}_{j+1} - 2~\rm{x}_i)~f~]^2 ~\sigma_y^2 .
\end{eqnarray}

Note that $\sigma_x$ and $\sigma_y$ are the rms noise in the LCP and RCP spectra, respectively, and that they were derived using line-free channels.

Thus, from equation~(\ref{q2B_eq}), 
\begin{equation}
\sigma_B = \frac{\sigma_q}{\sqrt{2}A_{\rm{F-F'}}}.
\end{equation}

\subsection{Monte Carlo Simulations}\label{montecarlo_subsec}

We performed Monte Carlo simulations to understand the proper implementation of both the $S$-curve and the cross correlation method, and to ensure that they have the same sensitivity. During these simulations we generated synthetic Zeeman $S$ spectra as produced by a strong $B$ field. In order to simulate data, we injected  repeatedly typical noise levels to the synthetic spectra and sampled them at typical channel spacings (here 0.164\kmss, see \S~\ref{obs_sec}). We subjected the ensemble of spectra to the same analysis procedures as the real data set, measuring B$_{||}$. The uncertainty for the individual $B_{||}$ field ``measurement'', $\sigma_{B_{||}}$, was derived from a formal minimum $\chi^{2}$ fit for the $S$-curve method and from error propagation considerations for the cross correlation method.

\begin{figure*}[!ht]
\plotone{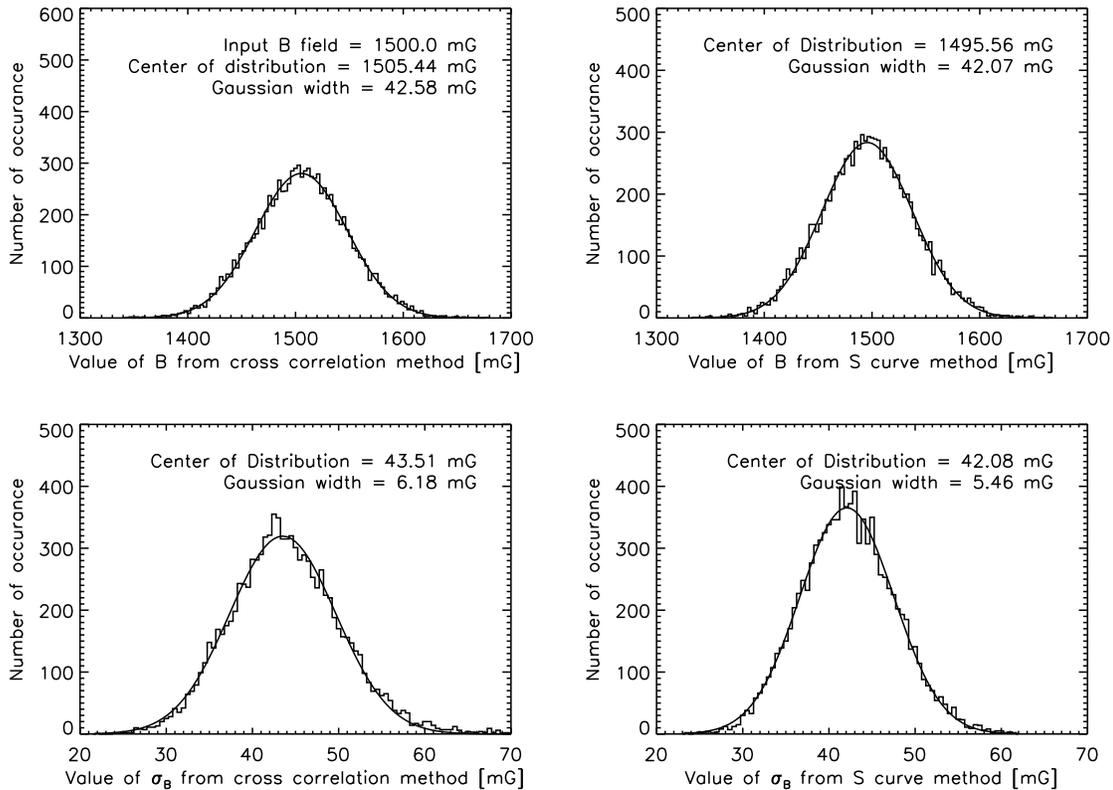}
\caption{Results of 10,000 Monte Carlo simulations of the $S$-curve and the cross correlation method. They were performed for synthetic data that consist of a Gaussian line with an amplitude of 1212 mJy and a line with of 0.79 \kms~(FWHM) plus an rms noise of 2.9 mJy (or a resolution of 0.16\kmss). The histograms show the distribution of ``measured'' magnetic fields $B$ and their associated uncertainties $\sigma_B$. Superimposed are Gaussian fits to the histograms and their parameters are displayed. Both methods are able to recover the applied $B$ field over the statistical range of simulations and both have similar uncertainties. Note that the parameters of the simulated data shown here are those of the GBT high-velocity feature at 1303~\kms~(see \S~\ref{polresults_highv_sec}).}
\label{montecarlofig}
\end{figure*}

As Figure~\ref{montecarlofig} shows, both methods can recover the input magnetic field strength and both have similar uncertainties. The uncertainty in the $B_{||}$ field estimation is expressed in two ways: a) the width of the histogram of the Gaussian distributed $B_{||}$ field measurements (top panel, Fig.~\ref{montecarlofig}) displays the intrinsic uncertainty in performing Zeeman measurements on a dataset with the characteristic noise and data qualities; b) the value for the center of the distribution of $\sigma_{B_{||}}$, the $B_{||}$ field uncertainties (bottom panel) as reported by the two Zeeman methods. Since the values for a) and b) coincide for both methods, we are confident that $i)$ the newly developed cross correlation technique is properly measuring the Zeeman splitting and $ii)$ our estimation of the uncertainties for both methods are accurate. With these simulations, we are also able to get a handle on the uncertainty of the uncertainty in the $B_{||}$ field estimate, $\sigma_{\sigma_{B_{||}}}$, indicated by the width of the histogram for the measured $B_{||}$ field uncertainties (bottom panel). Again, the values for $\sigma_{\sigma_{B_{||}}}$ coincide for both methods, as expected.

\bibliographystyle{apj}

\end{document}